\documentclass[lettersize,journal]{IEEEtran}
\usepackage{amsmath,amsfonts}
\usepackage{algorithmic}
\usepackage{algorithm}
\usepackage{array}
\usepackage[caption=false,font=normalsize,labelfont=sf,textfont=sf]{subfig}
\usepackage{textcomp}
\usepackage{stfloats}
\usepackage{url}
\usepackage{verbatim}
\usepackage{graphicx}
\usepackage{cite}
\usepackage{booktabs}
\usepackage{amsmath,amssymb,amsfonts}
\usepackage{algorithmic}
\usepackage{graphicx}
\usepackage{textcomp}
\usepackage{booktabs}
\usepackage{xcolor}
\usepackage{comment}
\usepackage{tikz}
\usepackage{orcidlink}
\usepackage{multirow}
\usetikzlibrary{matrix} 

\newcommand{\blackcircle}[1]{\tikz[baseline=(char.base)]{
            \node[shape=circle,fill=black,text=white,inner sep=1pt] (char) {#1};}}

\begin{document}

\title{TRACE: Unlocking Effective CXL Bandwidth via Lossless Compression and Precision Scaling}


\author{Rui Xie\orcidlink{0000-0003-3177-5071}, Asad Ul Haq\orcidlink{0009-0003-7975-0102}, Yunhua Fang\orcidlink{0009-0009-4718-8825}, Linsen Ma\orcidlink{0009-0000-8535-7911}, Zirak Burzin Engineer, Liu Liu\orcidlink{0000-0003-0792-8146}, Tong Zhang\orcidlink{0009-0009-8005-0043}

\thanks{Rui Xie, Asad Ul Haq, Yunhua Fang, Linsen Ma, Liu Liu and Tong Zhang are with Rensselaer Polytechnic Institute, Troy, NY, USA.}
\thanks{Zirak Burzin Engineer is with Wiseburn Da Vinci Science, El Segundo, CA, USA.} 
}



\maketitle

\begin{abstract}
LLM inference is increasingly limited by memory bandwidth, and the bottleneck worsens at long context as the KV cache grows. CXL memory adds capacity to offload weights and KV, but its link and device-side DDR bandwidth are far below HBM, so decoding stalls once traffic shifts to the CXL tier. Many CXL controllers are starting to add generic \emph{lossless} compression, yet applying commodity codecs directly to standard word-major LLM tensors is largely ineffective, especially for token-major KV streams. We propose TRACE (\textbf{T}raffic-\textbf{R}educed \textbf{A}rchitecture for \textbf{C}ompression and \textbf{E}lasticity), which preserves the unmodified CXL.mem interface but changes the device-internal representation. It stores tensors in a channel-major, disaggregated bit-plane layout, and applies a KV-specific transform before compression, converting mixed-field words into low-entropy plane streams that commodity codecs can compress. The same substrate enables precision-proportional fetch by reading only the required bit-planes. Across public LLMs, TRACE reduces BF16 weight footprint by 25.2\% and BF16 KV footprint by 46.9\% losslessly, with per-layer KV ratios peaking at 2.69$\times$. In trace-driven system modeling, once KV spills to CXL, GPT-OSS-120B-MXFP4 improves throughput at 128k tokens from 16.28 to 68.99 tok/s (4.24$\times$). DRAMSim3 shows up to 40.3\% lower DRAM access energy under plane-aligned fetch. 
A 7\,nm SystemVerilog implementation sustains 256\,GB/s device bandwidth. 
Relative to a CXL controller with generic inline lossless compression, TRACE only adds 7.2\% area, 4.7\% power, and 6.0\% load-to-use latency at 2\,GHz and 0.7\,V.

\end{abstract}

\vspace{-10pt}
\section{Introduction}\label{sec:introduction}
\IEEEPARstart{L}{arge} language model (LLM) inference has become a memory-traffic problem. GPUs keep adding compute, but decoding still stalls when the system cannot move model state fast enough. This gap shows up sharply once a substantial fraction of model state (KV and/or weights) spill beyond HBM and into a slower capacity tier. Compute Express Link (CXL) has therefore become the practical choice for expansion memory because it provides load-store access to DDR-class capacity at much lower cost per GB than HBM~\cite{sharma2023introduction,cxl101_cost,zhong2024managing,NVIDIA_H200,semianalysis_memory_wall}. The cost relief is real, but the bandwidth penalty is also real. A modern HBM stack delivers \(\sim\)1~TB/s-class bandwidth, while a CXL device sits behind a PCIe-class link and DDR channels with much lower sustained bandwidth~\cite{pcisig_pcie7_spec,jedec_lpddr6_standard}. Once the working set crosses the HBM boundary, the CXL tier can pin end-to-end throughput.

This leads to a first-principles framing. The CXL tier stalls decoding when the system must move too many bytes per token. A system can respond in two ways. It can (i) touch fewer data objects, or (ii) move fewer bytes per object. Recent algorithmic work has made strong progress on (i) by shrinking the active working set through quantized weights, selective expert execution, and page-level KV management~\cite{frantar2022gptq,lin2023awq,raposo2024mixture,liu2023deja,tang2024quest}. These techniques are effective and we view them as complementary. But they do not eliminate a key pressure point in long-context serving: the system still must keep and access a set of high-value KV pages at high precision, and each access moves full FP16/BF16/FP8 payloads over the device internal buses and the CXL link. In other words, page policies can reduce how many pages the system keeps, but they do not reduce the bytes moved for the pages the system must keep.

\begin{figure}[t]
    \centering
    \includegraphics[width=\linewidth]{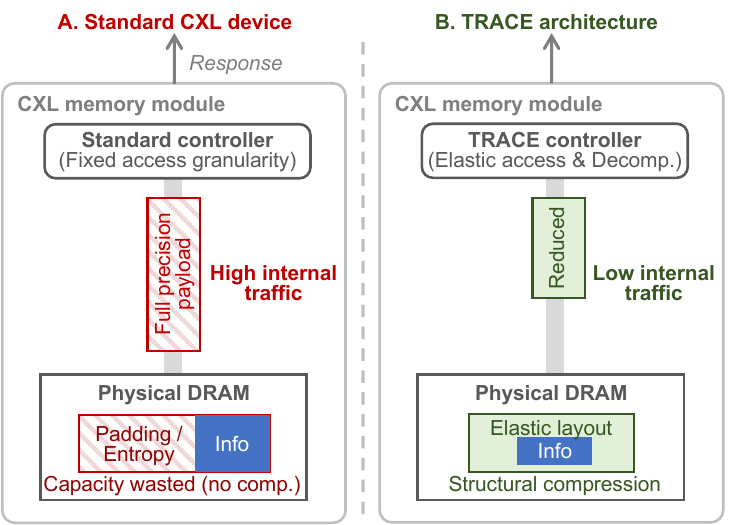}
    \caption{\textbf{The bandwidth and capacity gap in CXL memory.} (A) Standard devices enforce a rigid, word-based layout. This creates two inefficiencies: capacity waste (high entropy prevents compression) and bandwidth waste (fetching full words even when low precision is needed). (B) \textbf{TRACE} transforms the internal representation. By restructuring data into an elastic layout, it enables structural compression (saving KV capacity) and precision-proportional fetching (saving weight bandwidth), amplifying the effective capability of the CXL tier.}
    \label{fig:teaser}
\end{figure}

This leaves system designers with a practical dilemma. Aggressive KV compression via low-bit quantization or eviction can extend context length, but it can also discard information or introduce quantization error that hurts long-context quality. Keeping KV in BF16/FP16 preserves accuracy, but it pushes the CXL tier toward a bandwidth wall. The natural question is whether the system can keep high-precision KV \emph{exact} while paying a much lower bandwidth and capacity cost. This paper answers yes, but only if the device changes how it stores tensors internally.
Two obstacles block the obvious approaches on today's CXL devices (as shown in Fig.~\ref{fig:teaser}~(A)).

\noindent$\bullet$ \textbf{Issue 1: Direct lossless compression is ineffective under the standard layout.}
Applying fast codecs such as LZ4 or ZSTD directly to raw LLM tensors yields little or no reduction on KV cache and only modest reduction on weights (Table~\ref{tab:direct compression ratios}). The reason is structural. Standard floating-point storage interleaves sign, exponent, and mantissa bits at byte granularity, and KV is written token-by-token. Together they hide cross-token similarity and expose high-entropy byte streams to the compressor. As a result, the hardware sees data that is difficult to compress even when strong redundancy exists.

\noindent$\bullet$ \textbf{Issue 2: Fine-grained precision tiers do not automatically translate into proportional traffic savings.}
Modern runtimes increasingly explore more than a binary choice for KV pages (keep vs drop) by assigning multiple precision tiers to balance quality and capacity~\cite{tang2024quest,liu2024kivi}. However, a conventional CXL device stores tensors as fixed-width words and serves requests at cache-line granularity. This forces the device to move padded containers and full lines even when the runtime only needs fewer effective bits. Without a device-side representation that can physically skip unneeded bits, finer precision structure at the software level is hard to convert into a proportional reduction in bytes moved inside the device and over the link.

These two issues point to the same conclusion. The bottleneck is not a missing algorithm. The bottleneck is the device's internal representation. We are therefore forced to rethink how tensors are laid out inside the CXL tier.

We present \textbf{TRACE} (\textbf{T}raffic-\textbf{R}educed \textbf{A}rchitecture for \textbf{C}ompression and \textbf{E}lasticity), a transparent near-data architecture for CXL Type-3 memory shown in Fig.~\ref{fig:teaser}. TRACE keeps the CXL.mem interface unchanged and does not require application changes. Instead, it performs a device-internal layout transformation that exposes tensor structure to the memory controller.

\noindent$\bullet$ \textbf{Contribution 1: Make generic lossless compression effective for LLM state in the CXL tier.}
TRACE preserves the CXL.mem load/store interface but changes the device-internal representation to a channel-major, bit-plane layout, and applies a KV-specific transform  before compression. This converts token-major KV and mixed-field words into low-entropy plane streams that commodity codecs can compress. Across public LLMs, TRACE reduces BF16 weight footprint by 25.2\% and BF16 KV footprint by 46.9\% with no accuracy loss, with per-layer KV ratios peaking at 2.69$\times$. On GPT-OSS-120B-MXFP4, TRACE improves throughput at 128k tokens by 4.24$\times$ once KV spills to CXL.

\noindent$\bullet$ \textbf{Contribution 2: Turn multi-tier precision into physical DRAM-byte savings via plane-aligned fetch.}
TRACE exposes precision views as address aliases and serves them by fetching only the required bit-planes, so DRAM activations and bytes moved scale with the requested precision rather than a fixed word width. TRACE reduces weight-read DRAM access energy by up to 29.9\% (per-expert) and up to 40.3\% (per-head/per-neuron on OPT~30B), and reduces model-load latency by up to 30.0\%. A 7\,nm controller implementation preserves the unmodified CXL.mem interface and adds 7.2\% area, 4.7\% power, and 6.0\% load-to-use latency relative to a baseline with generic inline lossless compression.


\section{Background and Motivation}\label{sec:background}

\subsection{CXL Device As Extension}
CXL exposes DDR-class memory behind a CXL Type-3 device (CXL.mem) as byte-addressable capacity that the OS can map and manage like regular memory~\cite{sharma2023introduction}. In LLM deployments, accelerators keep the hot working set in HBM and place bulk state in a cheaper capacity tier, where CXL is a natural fit for weights and KV overflow~\cite{cxl101_cost,zhong2024managing}. Pools of such memory can be shared across hosts and typically cut effective cost per GB by about 50--55\% in modeled deployments~\cite{cxl101_cost}. 

The gap is bandwidth. A single HBM3E stack delivers roughly 1~TB/s-class bandwidth, whereas the CXL host link and the device-side DRAM sit far lower: PCIe~7.0~x16 targets 256~GB/s per direction, and DDR-class channels provide 51.2~GB/s at DDR5-6400 and 102.4~GB/s at DDR6-12800~\cite{semianalysis_memory_wall,pcisig_pcie7_spec,jedec_lpddr6_standard}. Current CXL devices inherit the conventional DRAM organization: tensors are stored as contiguous fixed-width words, and the controller moves full cache lines at that word width. Any wasted bits on the device DRAM bus therefore translate directly to extra DRAM beats, higher latency, and higher energy in the CXL tier.

This section motivates one target: raise effective bandwidth in the CXL tier under an unmodified CXL.mem interface. The most direct path is to move fewer bytes per accessed object. We next show why direct lossless compression fails to deliver that benefit under today’s layout.

\subsection{Lossless Compression Under Word Layout}\label{sec:lossless_motivation}

Long context inference increasingly becomes KV-dominated because the two memory terms scale differently. Model weights are a fixed, one-time footprint that can be prefetched and reused across tokens, while the KV cache grows linearly with both context length and batch size and must be accessed (read and updated) during decoding~\cite{liu2023deja}. As context and batch increase, KV therefore becomes the dominant consumer of capacity and memory bandwidth in inference serving.


Lossless compression is the cleanest way to reduce traffic without paying an accuracy tax. However, generic codecs are weak on raw LLM tensors stored in the standard word-based floating-point layout. We evaluate LZ4~\cite{lz4-link} and ZSTD~\cite{zstd-link} \textit{directly} on weights and KV for common models on the BookSum dataset. Table~\ref{tab:direct compression ratios} shows that under the conventional layout, LZ4 provides little savings and ZSTD yields only modest gains on weights and limited gains on KV.

\begin{table}[htbp]
\centering
\caption{Model weights and KV cache footprint reduction under direct lossless compression.}
\label{tab:direct compression ratios}
\resizebox{\columnwidth}{!}{%
\begin{tabular}{@{}lccccc@{}}
\toprule
\textbf{Codecs} & \textbf{LLaMA 3.1 8B} & \textbf{Gemma 2 2B} & \textbf{Mistral 7B} & \textbf{OPT 13B} & \textbf{Mixtral 8$\times$7B} \\ \midrule
\multicolumn{6}{c}{\textbf{Model Weights}} \\ \midrule
LZ4  & 0.0\%  & 11.5\% & 0.0\% & 0.0\% & 18.0\% \\
ZSTD & 20.6\% & 23.0\% & 17.3\% & 19.4\% & 21.3\% \\ \midrule
\multicolumn{6}{c}{\textbf{KV Cache on BookSum Dataset}} \\ \midrule
LZ4  & 0.0\% & 0.0\% & 0.0\% & 0.0\% & 0.0\% \\
ZSTD & 6.5\% & 2.9\% & 0.9\% & 2.0\% & 3.8\% \\ \bottomrule
\end{tabular}%
}
\end{table}

This behavior is not surprising once we look at what the codec actually sees. IEEE floating-point values interleave sign, exponent, and mantissa at byte granularity, so even small exponent variation quickly destroys byte-level repetition. KV further exacerbates this because it is produced token by token, so the write order fragments any cross-token regularity into many short, high-entropy streams. In other words, the conventional layout does not preserve the natural axes along which KV evolves smoothly, so a byte-oriented compressor has little opportunity to form long matches.

\begin{figure}[htbp]
    \centering
    \includegraphics[width=.9\linewidth]{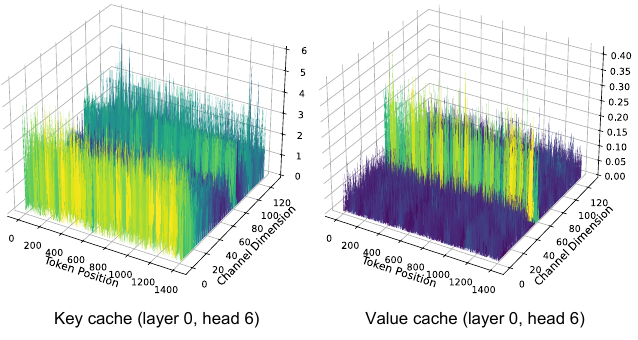}
    \caption{Activations are structurally smoother along channels. Visualization of Llama-3-8B KV cache on Booksum (layer 0, head 6) shows that values change less rapidly along the channel dimension (y-axis) than across tokens (x-axis), indicating latent structure that a token-major byte stream can obscure.}
    \label{fig:kv-visualize}
\end{figure}

Fig.~\ref{fig:kv-visualize} provides a clue about where the structure lives. In both the key and value caches, the surface changes less abruptly when moving along the channel dimension than when moving along token position. The patterns are coherent rather than noise-like, which indicates that KV has exploitable regularity, but that regularity becomes difficult to access when the system exposes KV as token-major byte streams.

The takeaway is that the tensors are not inherently incompressible. The bottleneck is representational. Under the conventional word layout, the system presents weight and KV tensors to the memory tier as short, mixed-field byte streams that obscure cross-element structure. This raises a central systems question for the rest of the paper: How can a CXL memory tier preserve and exploit this structure to reduce traffic, while keeping the CXL.mem interface and existing runtimes unchanged?

\subsection{Runtime Structure}\label{sec:runtime_structure}

Lossless compression reduces the \emph{cost per byte moved}. The remaining question is demand: what structure does the runtime expose in \emph{what} it reads and writes. This matters most for long-context decoding, where each token consults historical context through attention and updates KV every step~\cite{liu2023deja}. 

\smallskip
\noindent\textbf{KV is page-managed and long-tailed.}\quad
KV grows with context and is accessed every decoding step. Practical systems therefore manage KV as pages and make coarse keep/drop decisions based on runtime importance signals such as recency and attention mass~\cite{tang2024quest}. Table~\ref{tab: dynamic quantization perplexity} captures the typical outcome: purely dropping pages reduces traffic but can degrade quality, while retaining more pages at reduced precision can recover quality. The key runtime fact is that page importance is not binary, it is long-tailed. This naturally motivates finer-grained tiers of KV treatment at the page level, rather than a single keep-or-evict rule.

\begin{table}[htbp]
\centering
\caption{Perplexity for various page-level KV policies on LLaMA 3.1 8B and BookSum.}
\label{tab: dynamic quantization perplexity}
\resizebox{\columnwidth}{!}{%
\begin{tabular}{@{}ll@{}}
\toprule
\textbf{Method}                                                          & \textbf{Perplexity} \\ \midrule
Full KV Cache                                                            & 10.49               \\
Sliding Window (64 tokens)                                               & 14.33               \\
Quest (Top 5 pages in BF16)                                              & 12.49               \\
Dynamic Quant. (Top 5 in BF16, Next 3 in FP8, Next 2 in FP4)            & 11.87               \\
Dynamic Quant. (Top 5 in BF16, Next 5 in FP8)                            & 11.60               \\ \bottomrule
\end{tabular}%
}
\end{table}

\begin{figure}[htbp]
  \centering
  \includegraphics[width=.9\linewidth]{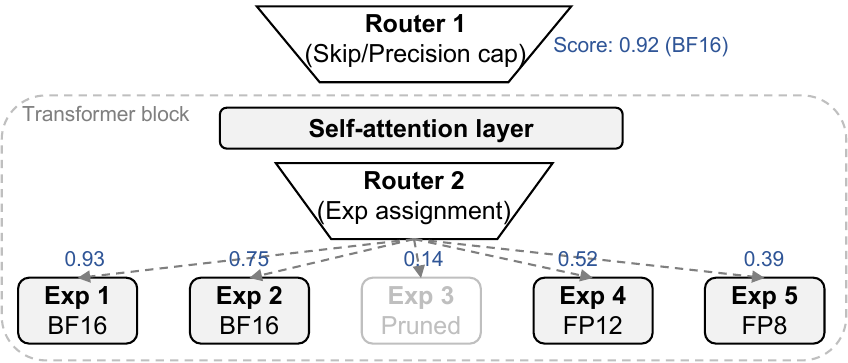}
  \caption{Dynamic weight precision with MoDE: a block-level cap (Router 1) and per-expert assignments (Router 2).}
  \label{fig.illustration MoDE dynamic quantization}
\end{figure}

\begin{figure}[htbp]
  \centering
  \includegraphics[width=\linewidth]{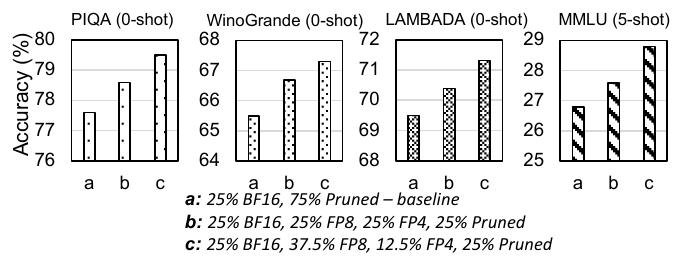}
  \caption{MoDE per-expert precision vs prune-only on LLaMA-MoE-3.5B for PIQA~\cite{bisk2020piqa}, WinoGrande~\cite{sakaguchi2021winogrande}, LAMBADA~\cite{paperno2016lambada}, and MMLU~\cite{hendrycks2020measuring}.}
  \label{fig.moe dynamic quantization}
\end{figure}

\smallskip
\noindent\textbf{Weights are block-read and often conditional.}\quad
Weight traffic is large and repeats every token. Two runtime facts matter. First, many deployments already use low-bit \emph{static} weight formats (e.g., INT8/INT4) to reduce footprint and bandwidth~\cite{frantar2022gptq,lin2023awq}. Second, conditional computation reduces the \emph{set of blocks} that execute per token, so only a subset of expert weights are read on that step~\cite{raposo2024mixture,liu2023deja}. Fig.~\ref{fig.illustration MoDE dynamic quantization} visualizes the block/expert granularity exposed by MoDE (Mixture-of-Depth-and-Expert) routing~\cite{raposo2024mixture}, and Fig.~\ref{fig.moe dynamic quantization} shows that controlling expert participation yields smooth resource-quality trade-offs.

\begin{figure}[htbp]
  \centering
  \includegraphics[width=.9\linewidth]{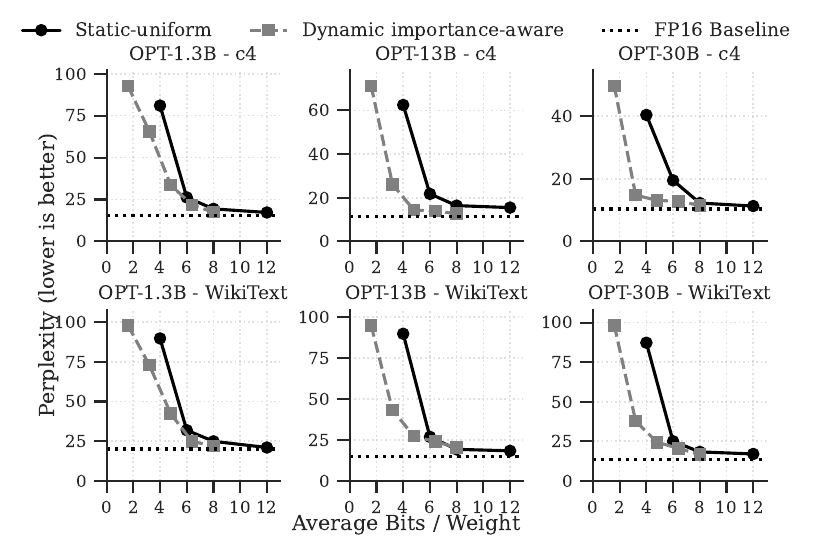}
  \caption{Perplexity vs average bits/weight on OPT. Per-head/per-neuron precision control outperforms static-uniform at the same bits. Lower is better.}
  \label{fig.perplexity}
\end{figure}

Even in dense transformers, importance is not uniform across heads and neurons. Fig.~\ref{fig.perplexity} shows that exploiting this heterogeneity can improve quality at the same average bit budget~\cite{liu2023deja}, which indicates structured non-uniformity in the weight-state that the runtime consumes.

\smallskip
\noindent\textbf{Compute already scales with precision.}\quad
Accelerators expose multiple native precision points, and public system specs report large throughput scaling as precision drops~\cite{NVIDIA_GB200_NVL72,NVIDIA_DGX_B200}. This makes a direct systems question unavoidable: can the memory tier co-scale bytes moved with the long-tailed KV importance and the structured weight demand that the runtime already exposes?

\section{Proposed Solution}\label{sec:design}

Section~\ref{sec:background} showed a clear mismatch in current systems. Modern inference algorithms use dynamic precision and have structured data like correlated KV channels. Standard memory devices ignore this. They store data in rigid blocks that mix up the bits, making them hard to compress. This forces the CXL tier to waste bandwidth on data that is redundant or not needed for the current math.

We propose \textbf{TRACE} (\textbf{T}raffic-\textbf{R}educed \textbf{A}rchitecture for \textbf{C}ompression and \textbf{E}lasticity), a near-data architecture to fix this. TRACE does not require new compression algorithms or changes to existing applications. Instead, it works as a transparent layer that changes how tensors are physically stored inside the CXL device.

\begin{figure}[htbp]
    \centering
    \includegraphics[width=.85\linewidth]{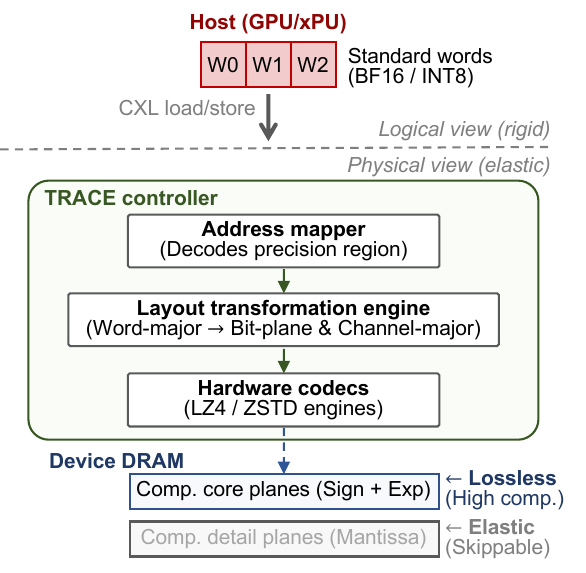}
    \caption{\textbf{TRACE overview.} The controller decouples the logical word-based interface from the physical bit-plane storage. An address mapper directs requests to optimized write paths: channel-major transposition for the dynamic KV cache and direct bit-plane encoding for static weights. During retrieval, the architecture enables elastic access: the controller selectively fetches specific compressed planes based on the host's precision requirement, skipping unnecessary data and applying optional on-device rounding.}
    \label{fig:overview}
\end{figure}

Fig.~\ref{fig:overview} shows the system architecture. TRACE separates the internal DRAM layout from the standard CXL.mem interface. This allows it to manage both static weights and dynamic KV caches using a single framework. The design relies on a core layout change called bit-plane disaggregation. This transformation enables two mechanisms to increase effective bandwidth:

\begin{itemize}
    \item \textbf{Structure-aware lossless compression:} By grouping similar bits together, such as exponents or aligned KV channels, TRACE creates smooth data streams that hardware codecs can easily compress. This increases the amount of data the link can carry.
    \item \textbf{Elastic precision access:} By storing precision levels in separate physical layers, TRACE makes the memory interface flexible. If the runtime needs less precision, the controller simply skips sending the detailed bit-planes. This saves bandwidth without changing the application logic.
\end{itemize}

\subsection{The Physical Substrate: Bit-Plane Disaggregation}
\label{sec:bitplane}

Standard memory devices enforce a rigid word-major layout. Regardless of the data type (BF16, INT8, or FP8), storing values as contiguous words physically interleaves bits of differing significance. This creates a high-entropy byte stream that obscures the spatial correlation of high-order bits (such as exponents) and forces the controller to perform fixed-granularity transfers. TRACE adopts bit-plane disaggregation as its fundamental physical substrate to resolve this. This transformation is a lossless transposition that groups bits by significance, effectively separating structural information from precision noise.

\smallskip
\noindent\textbf{Math notation.}\quad
We formalize this transformation as a matrix transposition over fixed-size memory blocks. Consider a block of $m$ values $\{x_0, \dots, x_{m-1}\}$, each encoded in $B$ bits. Let $\text{bit}(x_j, i)$ denote the $i$-th bit of the $j$-th value. In a conventional word-major layout, memory stores the logical bit-matrix $\mathbf{X}$ of size $m \times B$:
\begin{equation}
\mathbf{X}[j,i] = \text{bit}(x_j, i).
\label{eq:bitplane_x}
\end{equation}
TRACE physically stores the transpose of this matrix, $\mathbf{P} = \mathbf{X}^{\mathsf{T}}$. This reorganization results in $B$ distinct bit-planes stored as contiguous byte streams in DRAM. The $i$-th plane $P_i$ collects the $i$-th bit position across all $m$ values:
\begin{equation}
\mathbf{P} = 
\begin{bmatrix}
P_{B-1} \\
\vdots \\
P_0
\end{bmatrix}
=
\begin{bmatrix}
\text{bit}(x_0, B-1) & \cdots & \text{bit}(x_{m-1}, B-1) \\
\vdots & \ddots & \vdots \\
\text{bit}(x_0, 0) & \cdots & \text{bit}(x_{m-1}, 0)
\end{bmatrix}
\label{eq:bitplane_p}
\end{equation}

\begin{figure}[htbp]
    \centering
    \includegraphics[width=\linewidth]{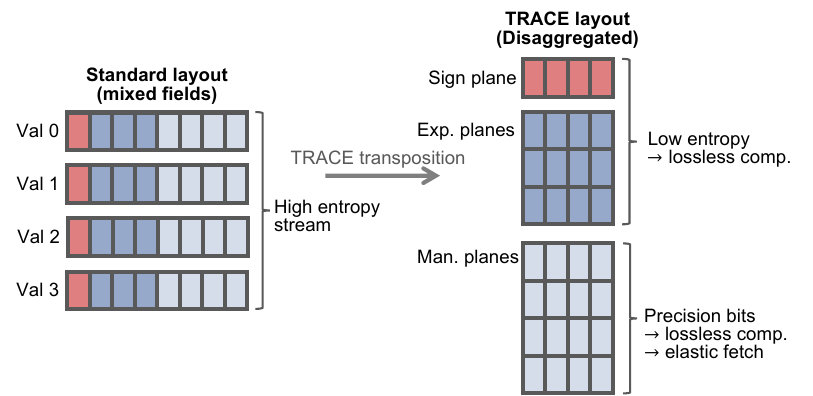}
    \caption{\textbf{Bit-plane disaggregation as a physical filter.} (A) The standard word-major layout interleaves fields, creating a high-entropy stream that obscures structure. (B) TRACE transposes data to physically group bits by significance. This exposes \textbf{low-entropy streams} (Sign/Exponent) for efficient lossless compression and isolates \textbf{precision details} (Mantissa) to enable elastic, precision-proportional fetching.}
    \label{fig:bitplane_transform}
\end{figure}

\smallskip
\noindent\textbf{Structural decoupling.}\quad
As illustrated in Fig.~\ref{fig:bitplane_transform}, this transposition acts as a physical filter that partitions data into two functional streams based on bit significance:
\begin{itemize}
    \item \textbf{Compressible core (Sign \& Exponent):} High-order planes (e.g., BF16 exponents) typically exhibit long runs of identical values due to the local magnitude stability of tensors. Grouping these bits creates low-entropy streams ideal for hardware compression.
    \item \textbf{Elastic detail (Mantissa):} Low-order planes contain fine-grained precision information, which often behaves like high-entropy noise. Physically isolating these layers allows the controller to skip reading them when the runtime requests lower precision.
\end{itemize}

\smallskip
\noindent\textbf{Line-rate implementation.}\quad
TRACE performs the transposition on fixed-size blocks (e.g., 4\,KB) aligned to DRAM rows. The controller uses an internal SRAM staging buffer to coalesce incoming cache-line writes and stream the transpose into the codec/DRAM-commit pipeline. In steady state, transposition is fully overlapped with buffering and compression and does not extend the critical path. Backpressure occurs only when sustained write demand exceeds the device drain rate beyond the available buffer capacity.

\subsection{Mechanism I: Structure-Aware Lossless Compression}\label{sec:compression}

The bit-plane substrate provides the physical foundation for compression, but it is not sufficient on its own for all data types. To maximize bandwidth, the system must address the specific structural properties of the dynamic KV cache and the static model weights separately.

\smallskip
\noindent\textbf{Pipeline for KV cache.}\quad
The primary barrier to compressing KV data is its arrival order. Host systems issue updates in a token-major format, placing values from different embedding channels into adjacent memory addresses. As analyzed in Section~\ref{sec:lossless_motivation}, adjacent channels often possess disparate scales, creating a high-entropy stream that resists standard compression. Therefore, Fig.~\ref{fig:kv-visualize} reveals the solution: values within the \textit{same} channel exhibit high temporal stability across sequential tokens.

\begin{figure}[htbp]
    \centering
    \includegraphics[width=\linewidth]{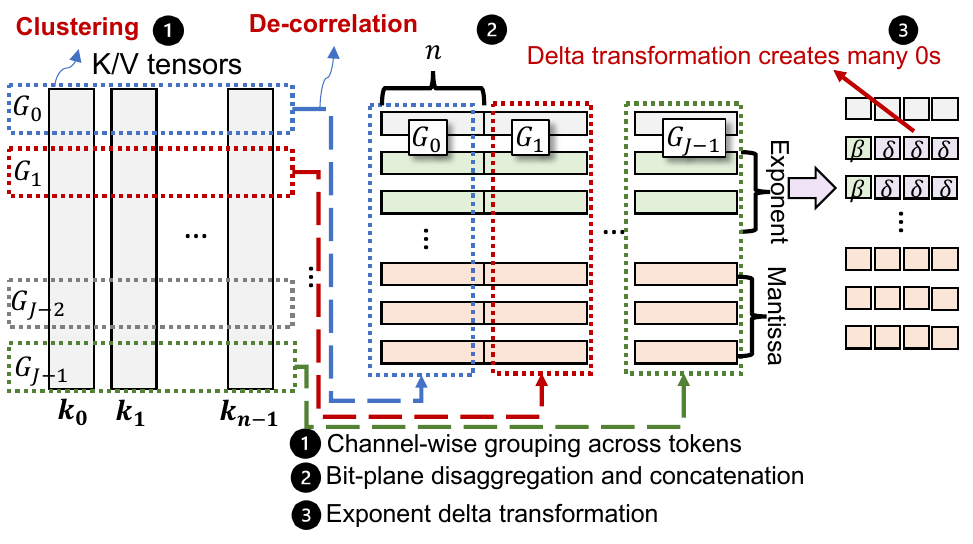}
    \caption{\textbf{Cross-token KV clustering.} TRACE buffers token-major updates and transposes them into channel-major blocks. High-correlation channel groups are then normalized via exponent deltas and split into bit-planes to maximize compression ratios by creating long runs of zeros in the MSBs.}
    \label{fig:cross_token_kv}
\end{figure}

To exploit this stability, TRACE applies a transformation chain that aligns the physical layout with the logical correlation of the data.
First, the controller resolves arrival fragmentation by buffering a window of $n$ tokens and transposing token-major updates into channel-major groups. For a channel $j$, we define the group $G_j$ as:
\begin{equation}
G_{j} = \{ k_{t, j} \mid t = 0, \dots, n-1 \}.
\end{equation}
This converts the stream from disparate vectors into stable per-channel time series (Step \blackcircle{1} in Fig.~\ref{fig:cross_token_kv}). The transposition is performed in an SRAM staging buffer that decouples host stores from internal packing. Backpressure is only possible when sustained KV write bandwidth exceeds the device drain rate for longer than the buffer can absorb, and the required buffer size scales with the runtime KV page budget. Concretely, buffering $n$ tokens for one writable KV stream/page requires
\begin{equation}
S_{\mathrm{buf}} = n \cdot C \cdot b + S_{\mathrm{ovhd}},
\end{equation}
where $C$ is the number of channels in the page, $b$ is bytes per element (2 for BF16), and $S_{\mathrm{ovhd}}$ is a small header/alignment overhead. Total SRAM is $S_{\mathrm{tot}} = N_{\mathrm{streams}} \cdot S_{\mathrm{buf}}$ under a fixed page budget.

Second, grouping alone is insufficient if the floating-point representation remains noisy. To reduce entropy within each group $G_j$, the controller selects a base exponent $\beta_j$ and transforms each element into a delta:
\begin{equation}
\delta_{t, j} = \text{Exponent}(k_{t, j}) - \beta_{j}.
\end{equation}
Since activation magnitudes are stable within a channel, these deltas are typically small (Step \blackcircle{2}).

Finally, the deltas are mapped to the bit-plane substrate (Step \blackcircle{3}). Small deltas induce sparse high-order planes (long runs of zeros), which improves the effectiveness of generic codecs.

\smallskip
\noindent\textbf{Direct application to weights.}\quad
Static model weights require a different approach. Unlike the KV cache, weights are not generated at runtime, so they do not require buffering or delta normalization. We apply lossless compression directly to the bit-planes of the weight tensors. This mechanism is effective even for quantized models. For example, if the host uses INT8 weights, the physical bit-planes of that format still contain structural redundancy, such as long runs of zeros in the most significant bits of outlier-heavy distributions. TRACE compresses these planes individually, creating a bandwidth multiplier that is orthogonal to the algorithmic quantization chosen by the user.

\smallskip
\noindent\textbf{Write path and codec integration.}\quad
On the KV write path, incoming cache-line stores are first coalesced into 4\,KB blocks in the SRAM staging buffer, then transformed (cross-token transposition + exponent-delta) and bit-plane packed, and finally compressed by a generic inline codec before DRAM commit. On the read path (KV or weights), the controller fetches the required compressed planes, decompresses them, and reconstructs standard cache lines for return. TRACE reuses commodity codecs (e.g., LZ4 for latency-sensitive updates and ZSTD where decode can be amortized); the gain comes from changing the codec input from mixed-field word streams to low-entropy plane streams rather than relying on a specialized compressor.

\subsection{Mechanism II: Elastic Precision Access}\label{sec:elasticity}

Beyond compression, TRACE transforms the rigid memory interface into an elastic one. As established in Section~\ref{sec:bitplane}, the device physically stores all tensors as independent bit-planes in a compressed format. TRACE exploits this unified abstraction to enable precision-proportional fetching through a linear pipeline of address translation, row filtering, and data reconstruction. 
TRACE is complementary to software techniques such as KV quantization and eviction: it reduces traffic for the pages the runtime still chooses to keep. While our evaluation (Section~\ref{sec:experiment}) focuses on transformer KV, the bit-plane substrate and the lossless path apply directly to other static tensors as well (i.e., weights in alternative architectures).

\begin{figure}[htbp]
  \centering
  \includegraphics[width=\linewidth]{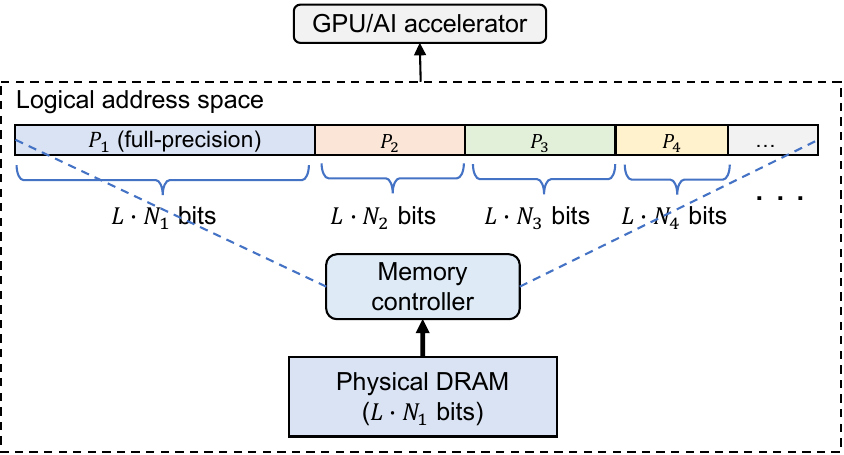}
  \caption{\textbf{Precision-partitioned logical address space.} TRACE maps multiple logical regions ($P_1 \dots P_k$) to the same physical data. Region $P_1$ represents the full-precision view ($L \cdot N_1$ bits), while regions $P_2 \dots P_k$ represent reduced-precision aliases ($L \cdot N_i$ bits). The host selects the desired precision via pointer arithmetic.}
  \label{fig:memory_virtualization}
\end{figure}

\smallskip
\noindent\textbf{Precision-partitioned addressing.}\quad
TRACE exposes elasticity via address aliasing without changing the CXL.mem protocol. Let $L$ be the number of elements in a tensor (or KV block) and $N_1$ the full bit-width (e.g., 16). The device physically stores $L\!\cdot\!N_1$ bits as shared bit-planes, while the driver exposes multiple virtual-address aliases $P_1,\dots,P_k$ that map to the same physical data (Fig.~\ref{fig:memory_virtualization}): $P_1$ is the full-precision view and each $P_i$ for $i>1$ is a reduced-precision view spanning $L\!\cdot\!N_i$ bits. The accessed alias determines which planes the controller returns, so load/store semantics and cache-line transfers are unchanged and no sideband signaling is needed. The lossless path is transparent once memory is allocated on a TRACE-backed region via the driver/allocator, while reduced-precision views are opt-in and require the runtime/allocator to select the corresponding alias pointer per tensor/page policy. Unused aliases need not be mapped (or can be provided via a small remappable alias window), and because all views alias the same planes, exposing additional views incurs no extra device DRAM capacity.

\smallskip
\noindent\textbf{Selective retrieval and guard planes.}\quad
When a read request hits an alias region $P_i$, the controller translates the logical intent into a specific set of physical rows. If region $P_i$ corresponds to a format with $N_i$ bits ($1+r_E+r_M$), TRACE generates a mask to retrieve exactly the following set of compressed planes $\mathcal{S}_{req}$:
\begin{equation}
\mathcal{S}_{req} = \{P^{\mathrm{sgn}}\} \cup \{P^{\mathrm{exp}}_{E-r_E, \dots, E-1}\} \cup \{P^{\mathrm{man}}_{M-r_M, \dots, M-1}\}
\end{equation}
Plane selection follows a fixed mapping for each target format: the controller always returns the sign plane and the most significant exponent/mantissa planes implied by $(r_E,r_M)$. It does not inspect per-element values to decide which planes to fetch.
To locate these variable-length compressed blocks without incurring double DRAM accesses, the controller queries an on-chip Plane Index (detailed in Section~\ref{sec:microarch}) to resolve physical offsets in a single cycle.
The DRAM scheduler uses this set to filter Row Access Strobe (RAS) commands. Rows containing the compressed lower-order mantissa planes are physically skipped.

To mitigate quantization error caused by simple truncation, TRACE supports on-device rounding. When an alias region is configured for high fidelity, the controller retrieves a defined number of guard planes ($d_E, d_M$) beyond the requested precision cut-off. The total fetch expands to $1 + (r_E + d_E) + (r_M + d_M)$ planes. The controller uses these extra bits, which effectively act as the guard and round bits in standard floating-point arithmetic, to perform round-to-nearest logic on-chip before the data is serialized. This mechanism significantly improves numerical accuracy for sensitive layers while incurring negligible bandwidth overhead, as the guard planes themselves are transmitted in their compressed form. This on-device rounding is particularly effective for preserving outlier values in activation channels, which are known to be sensitive to truncation errors.

\begin{figure}[htbp]
  \centering
  \includegraphics[width=\linewidth]{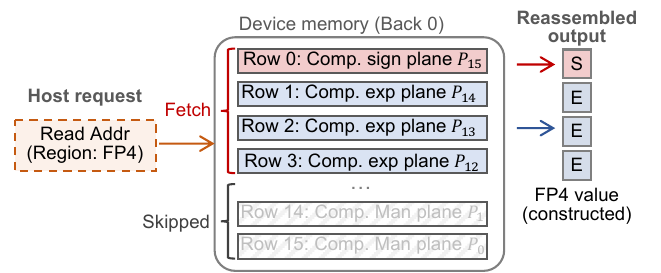} 
  \caption{\textbf{Elastic precision via selective row activation.} The logical request (left) drives the physical retrieval process (center). For a low-precision request, the DRAM controller physically activates only the rows containing the required \textbf{compressed bit-planes} (Sign/Exponent). Rows containing compressed LSB planes remain dormant, ensuring that bandwidth consumption scales linearly with the requested precision.}
  \label{fig:elasticity}
\end{figure}

\noindent\textbf{Reconstruction pipeline.}\quad
Since both weights and KV cache are stored as compressed bit-planes, the retrieval path utilizes a shared decompression engine. However, the final structural restoration differs based on the data type. We formalize the reconstruction pipeline as a composite function mapping the fetched compressed planes $\mathcal{S}_{req}$ to the host-visible tensor $\mathbf{X}_{out}$:
\begin{equation}
\mathbf{X}_{out} = \left( \mathcal{T}^{-1} \circ \mathcal{R} \circ \mathcal{D} \right) (\mathcal{S}_{req})
\end{equation}
This pipeline applies three operators in sequence:
\begin{enumerate}
    \item $\mathcal{D}$ (\textbf{decompression}): Standard codec decoding (e.g., LZ4/ZSTD) to recover raw bit-planes.
    \item $\mathcal{R}$ (\textbf{arithmetic reconstruction}): Reassembles numerical values from bit-planes. This operator handles precision adjustment: it applies the guard-plane rounding logic and zero-pads any missing LSB planes to restore valid 16-bit containers.
    \item $\mathcal{T}^{-1}$ (\textbf{inverse topology}): Restores the spatial layout expected by the host kernel. This operator is context-dependent:
    \begin{equation}
        \mathcal{T}^{-1}(\mathbf{x}) = 
        \begin{cases} 
            \mathbf{x} & \text{if Weight} \\
            \text{Transpose}(\mathbf{x}) & \text{if KV Cache}
        \end{cases}
    \end{equation}
\end{enumerate}
This abstraction ensures architectural transparency. This inverse shuffle is implemented via a hardware network in the read path, ensuring the host receives standard cache lines without software intervention. The host consistently receives standard, linearly addressed tensors, whether fully lossless or precision-adapted, while the underlying physical link carries only the entropy-reduced payload.

\subsection{Controller Microarchitecture}\label{sec:microarch}

\begin{figure}[htbp]
    \centering
    \includegraphics[width=\linewidth]{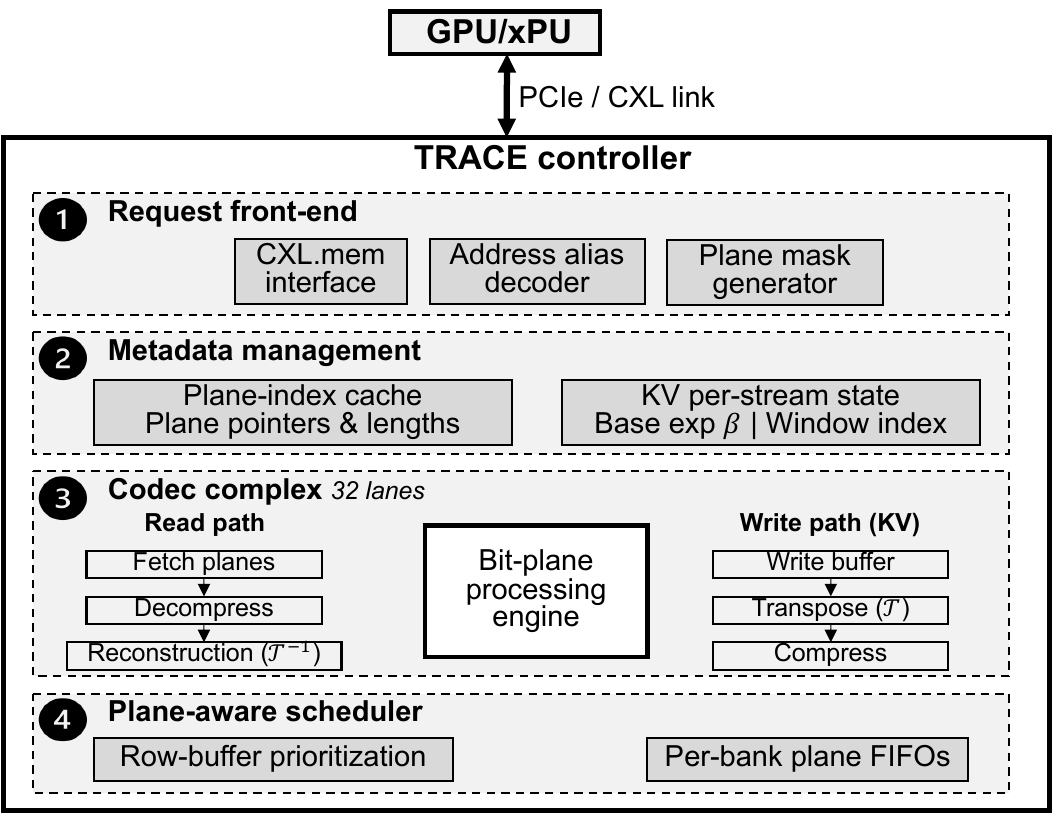}
    \caption{\textbf{TRACE Controller Microarchitecture.} The design features a four-stage pipeline: (1) The Request front-end translates logical aliases into physical plane masks; (2) Metadata management caches pointers in on-chip SRAM to avoid DRAM round-trips; (3) The Codec complex handles channel-major transposition and decompression at line rate; and (4) The Plane-aware scheduler optimizes row-buffer locality for bit-plane bursts.}
    \label{fig:microarchitecture}
\end{figure}

TRACE is implemented as a pipelined extension to a conventional CXL Type-3 (CXL.mem) controller. Fig.~\ref{fig:microarchitecture} shows the microarchitecture. Externally, the device preserves the standard cache-line-granularity CXL.mem interface and exposes the same host-visible tensor views. Internally, it stores data in a bit-plane substrate and makes two operations explicit in the controller: (i) translating a host request into a bit-plane selection (for precision-scalable fetch) and (ii) servicing that selection with plane-aligned DRAM transfers plus (de)compression.

\smallskip
\noindent\textbf{Request front-end.}\quad
The front-end accepts CXL.mem load/store commands and decodes whether the address targets the base region (full-precision, lossless) or an alias region (a reduced-precision view). The alias decoder maps the requested view into a \emph{plane mask} that specifies exactly which physical planes are required for the response. This plane mask is carried with the request so that downstream stages never fetch unused planes.

\smallskip
\noindent\textbf{Metadata management.}\quad
TRACE stores planes as variable-length compressed streams, so a request to a logical 4\,KB block must be translated into (i) the plane-bundle base pointer and (ii) per-plane compressed lengths, plus codec/bypass flags. The complete plane index is stored in device DRAM as a reserved metadata region, with one compact entry per 4\,KB block (64\,B per block, 1.56\% capacity overhead in our implementation). The controller caches a subset of index entries on-chip to avoid a DRAM round-trip on the common path; on a cache miss, it issues one additional DRAM read to fetch the index entry before issuing the data-plane reads (no speculative plane fetch or reread of data planes). For KV, the controller also maintains constant-size per-stream state (e.g., base exponent and window index) required by the KV transform and its inverse.

\smallskip
\noindent\textbf{Codec complex.}\quad
The codec complex is a set of parallel lanes (e.g., 32 lanes in Fig.~\ref{fig:microarchitecture}) organized around a bit-plane processing engine.
On the \emph{read path} (weights or KV reads), each lane (i) fetches the selected planes, (ii) decompresses them, and (iii) reconstructs the host-expected layout via $\mathcal{T}^{-1}$ before returning data on the CXL link.
On the \emph{KV write path}, incoming KV updates are buffered and transformed into the device-internal layout via $\mathcal{T}$, then compressed and committed to DRAM. The write buffer decouples host write timing from internal transposition and compression.

\smallskip
\noindent\textbf{Plane-aware scheduler.}\quad
TRACE schedules DRAM at plane granularity rather than word granularity. Requests are organized into per-bank plane FIFOs so that the controller can issue bursts that keep accesses within the same plane stripe, improving locality for plane-aligned reads. The scheduler applies row-buffer prioritization within each bank to reduce activate/precharge churn when serving multi-plane requests.

\smallskip
\noindent\textbf{Bypass and correctness invariants.}\quad
When a request targets an uncompressed region or a block that is stored incompressibly, the controller bypasses the codec and serves the corresponding planes directly. For any host-visible view, TRACE returns identical values to a baseline device serving the same view; the only difference is which internal planes are activated and how bytes are arranged on the device side.
Section~\ref{sec:ppa_latency} quantifies the added controller cost (area/power) and the exposed load-to-use latency under metadata hits/misses, compressible/incompressible blocks, and the bypass case.

\section{Evaluation}\label{sec:experiment}


\subsection{Evaluation Overview}
\label{sec:evaluation overview}
We evaluate TRACE by linking its two mechanisms to end-to-end throughput, device-local DRAM behavior, and hardware feasibility. Mechanism~I improves lossless compressibility by changing the device-internal representation (Sections~\ref{sec:bitplane} and~\ref{sec:compression}), and Mechanism~II enables precision-proportional fetch via plane-aligned access (Section~\ref{sec:elasticity}).

\smallskip
\noindent\textbf{Baselines.}\quad
All experiments use the same three CXL Type-3 baselines under identical traces and host-visible semantics (Table~\ref{tab:baseline_parity}): \textit{CXL-Plain} (word-major, no compression), \textit{CXL-GComp} (word-major + 4\,KB inline lossless compression with indexing and bypass), and \textit{TRACE} (bit-plane layout + KV preprocessing before the same codec, with optional plane-aligned fetch for alias views).

\begin{table}[htbp]
\centering
\caption{Evaluation baseline.}
\label{tab:baseline_parity}
\footnotesize
\setlength{\tabcolsep}{3.5pt}
\begin{tabular}{lccc}
\toprule
\textbf{Item} & \textbf{Plain} & \textbf{GComp} & \textbf{TRACE} \\
\midrule
CXL.mem cache-line interface & \checkmark & \checkmark & \checkmark \\
Device DRAM layout & word & word & bit-plane \\
4\,KB block codec + index cache + bypass & -- & \checkmark & \checkmark \\
KV cross-token transform & -- & -- & \checkmark \\
Plane-aligned fetch (alias views) & -- & -- & \checkmark \\
\bottomrule
\end{tabular}
\vspace{-2mm}
\end{table}

We report (i) trace-driven throughput modeling (Section~\ref{sec:trace-driven system modeling}), (ii) compression efficiency by layer and bit-plane (Section~\ref{sec:compression_efficiency}), (iii) DRAMSim3 energy/latency under plane-aligned fetch (Section~\ref{sec:DRAM access efficiency}), and (iv) RTL PPA and load-to-use timing (Section~\ref{sec:ppa_latency}).





\subsection{Trace-Driven System Modeling}
\label{sec:trace-driven system modeling}

We model decoding throughput with first-order bandwidth accounting in Python. For each setting, we compute per-token traffic on the CXL link and on the device-side DDR channels, then convert each to a tok/s ceiling by dividing the corresponding bandwidth by bytes-per-token and taking the bottleneck. 
The model is intended to isolate how bytes-per-token changes translate to throughput under fixed link/DDR bandwidth budgets.

Per-token traffic is decomposed into weight reads plus KV reads and writes. KV bytes are derived from the model shape (layers, KV heads, head dimension, and element size): each generated token appends one KV entry, and historical KV reads are modeled as a fixed fraction $f_{rd}$ of the context per step (e.g., 0.2), so KV read bytes scale linearly with sequence length. We treat $f_{rd}$ as a modeling parameter and use the same value across all baselines. The goal is to isolate bytes-per-token differences rather than predict absolute latency/queuing effects. Weight bytes are the per-token weight read volume implied by the model (including conditional execution when applicable).

Device-side DDR bytes are parameterized by measured 4\,KB-block footprints under each baseline’s device pipeline. We sample representative weight and KV blocks, apply the corresponding transform and 4\,KB block compression path, and use the resulting average compressed sizes to account for the stored bytes served by DDR (Section~\ref{sec:compression_efficiency}). HBM capacity is partitioned between weights and KV. Hits are approximated by capacity ratios under this fixed partition, and only the overflow portion is counted as CXL link/device traffic. We assume a single-GPU + CXL Type-3 system with a 512\,GB/s per-direction CXL link and 256\,GB/s device DDR bandwidth.

\smallskip
\noindent\textbf{Common case: weights fit in HBM; KV spill dominates.}\quad
We first study the regime where weights remain resident in HBM and only KV spills to the CXL tier. This matches a common deployment choice: weights incur a fixed per-token read cost, so systems prioritize keeping them on HBM and use remaining HBM capacity to cache the hottest KV pages. As context length (and concurrency) grows, KV expands and an increasing fraction of KV reads are served from CXL, so throughput becomes KV-traffic limited.

\begin{figure}[htbp]
    \centering
    \includegraphics[width=\linewidth]{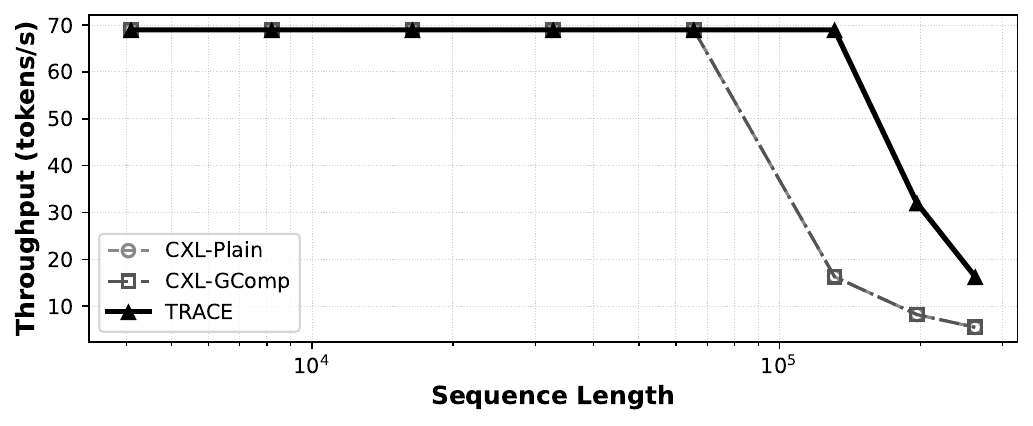}
    \caption{\textbf{Decoding throughput vs.\ context (GPT-OSS-120B-MXFP4; $\sim$60~GB weights fit in HBM).} All designs overlap up to 64k. Once KV spills to CXL (128k+), CXL-GComp overlaps CXL-Plain, while TRACE sustains higher throughput by reducing KV bytes moved under a KV-aware representation.}
    \label{fig:gptoss120b_mxfp4_seqlen}
\end{figure}

Fig.~\ref{fig:gptoss120b_mxfp4_seqlen} evaluates GPT-OSS-120B-MXFP4~\cite{agarwal2025gpt} (120B at 4-bit; $\sim$60~GB weights, fits in 76~GB usable HBM). All designs overlap up to 64k tokens at 68.99~tok/s, indicating CXL is not yet on the critical path. Once context reaches 128k and KV spill forces CXL reads, CXL-Plain drops to 16.28~tok/s and CXL-GComp remains essentially identical, while TRACE sustains 68.99~tok/s (4.24$\times$). TRACE continues to lead at longer contexts: 32.03 vs.\ $\sim$8.21~tok/s at 196k (3.90$\times$) and 16.28 vs.\ 5.49~tok/s at 256k (2.97$\times$). These trends match generic compression being ineffective on the token-major KV stream under a word-major layout, while TRACE’s KV-aware transformation reduces bytes moved.

\begin{figure}[htbp]
    \centering
    \includegraphics[width=\linewidth]{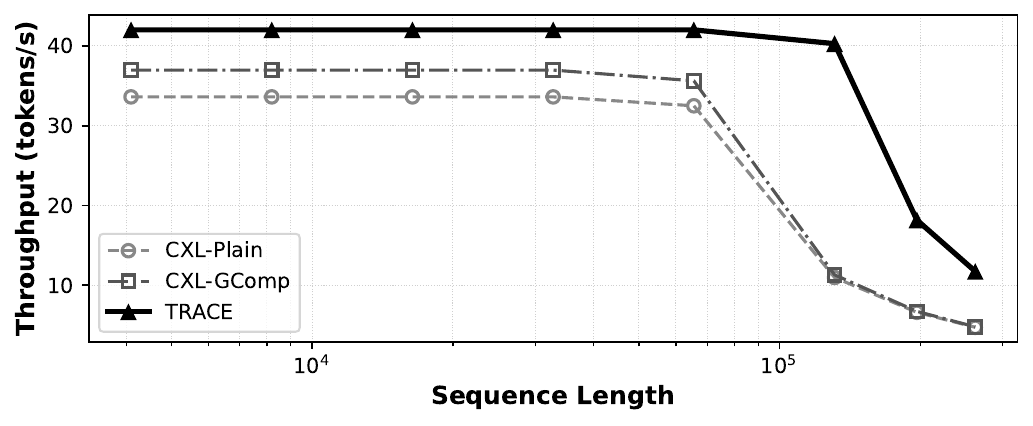}
    \caption{\textbf{Decoding throughput vs.\ context (GPT-OSS-120B; weights spill to CXL, \(\alpha=0.8\)).} TRACE improves short-context throughput under weight spill and remains higher at long contexts where KV spill dominates.}
    \label{fig:gptoss120b_seqlen}
\end{figure}

\begin{figure}[htbp]
    \centering
    \includegraphics[width=\linewidth]{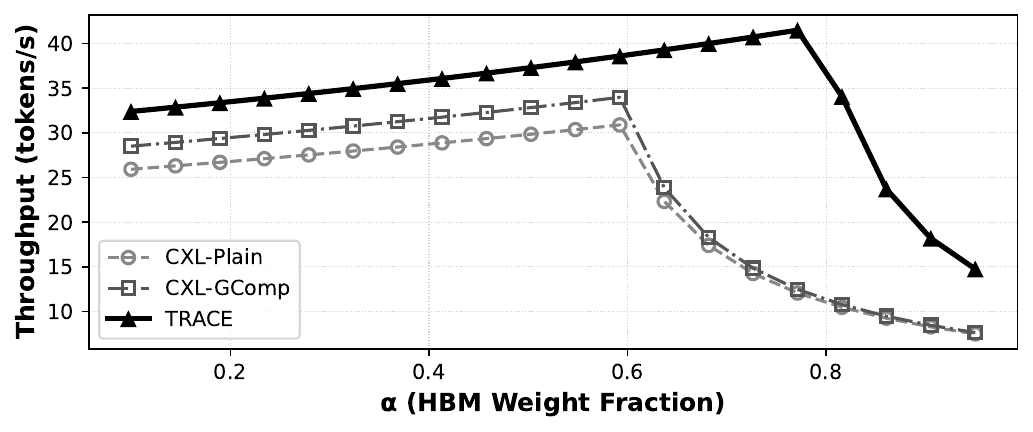}
    \caption{\textbf{Throughput sensitivity to HBM partitioning (\(\alpha\)) under weight spill (GPT-OSS-120B).} Throughput is unimodal in \(\alpha\): increasing \(\alpha\) reduces weight spill until KV spill dominates; TRACE raises the peak and shifts it to a larger \(\alpha\) by reducing effective bytes moved for spilled state.}
    \label{fig:gptoss120b_alpha}
\end{figure}

\smallskip
\noindent\textbf{Less-provisioned case: weights also spill to CXL.}\quad
We next consider a configuration where usable HBM cannot hold all weights, so both weights and KV can spill to CXL during decoding. To define the HBM split, we reserve an $\alpha\in(0,1)$ fraction of usable HBM for weights and the remaining $1-\alpha$ for the hot KV working set:
\begin{equation}
H_{w} = \alpha \cdot H_{\mathrm{user}}, \qquad
H_{kv} = (1 - \alpha) \cdot H_{\mathrm{user}}.
\end{equation}
Overflow weights and KV pages are served from CXL. We fix $\alpha=0.8$ in Fig.~\ref{fig:gptoss120b_seqlen} (weight-priority partition) and sweep $\alpha$ in Fig.~\ref{fig:gptoss120b_alpha} to show the trade-off.

Fig.~\ref{fig:gptoss120b_seqlen} evaluates GPT-OSS-120B~\cite{agarwal2025gpt} (BF16 $\sim$240~GB weights $>$ 76~GB usable HBM; $\alpha=0.8$). The curves separate already at 4k because weights must be fetched from CXL: CXL-Plain is 33.61~tok/s, CXL-GComp is 36.97~tok/s, and TRACE is 42.02~tok/s. CXL-GComp exceeds CXL-Plain because weights have non-zero lossless compressibility even under word-major layout (Table~\ref{tab:direct compression ratios}). TRACE is higher because it improves compressibility for both weights and KV. As context grows, all designs show a throughput cliff as KV traffic scales with sequence length and more KV reads spill to CXL. TRACE remains higher in the long-context regime: at 128k it sustains 40.29~tok/s while the baselines drop to 10.97--11.30~tok/s ($\sim$3.6$\times$). At a $\sim$40~tok/s target, TRACE reaches 128k whereas the baselines reach similar throughput only up to 64k ($\sim$2.00$\times$ longer context).

Fig.~\ref{fig:gptoss120b_alpha} sweeps $\alpha$ for GPT-OSS-120B in BF16. Throughput is unimodal because $\alpha$ trades HBM between weights and KV: small $\alpha$ causes heavy weight spill and low throughput, while large $\alpha$ starves the hot KV budget and increases KV spill. Numerically, CXL-Plain peaks at 30.89~tok/s at $\alpha=0.592$ (25.92~tok/s at $\alpha=0.10$) and drops to 7.47~tok/s at $\alpha=0.95$. CXL-GComp peaks at 33.98~tok/s at $\alpha=0.592$ (28.51~tok/s at $\alpha=0.10$), showing that generic compression mainly helps the weight-spill side but does not prevent the KV-driven drop as KV budget shrinks. TRACE reaches 41.51~tok/s at $\alpha=0.771$ (32.40~tok/s at $\alpha=0.10$) and stays higher over a wider range, consistent with reducing effective bytes moved when spilled state is served from CXL.

\subsection{Lossless Compression Efficiency}
\label{sec:compression_efficiency}

This subsection quantifies where TRACE’s lossless gains come from under a fixed 4\,KB block granularity and commodity codecs. We report compression ratio $S_{\mathrm{orig}}/S_{\mathrm{comp}}\ge 1$ (larger is better). Unless noted, all results use 4\,KB blocks with LZ4/ZSTD. TRACE always includes the bit-plane layout (Section~\ref{sec:bitplane}); for KV it additionally applies Mechanism~I (cross-token channel grouping + exponent-delta de-correlation, Section~\ref{sec:compression}).

\begin{figure*}[htbp]
	\centering
    \includegraphics[width=\linewidth]{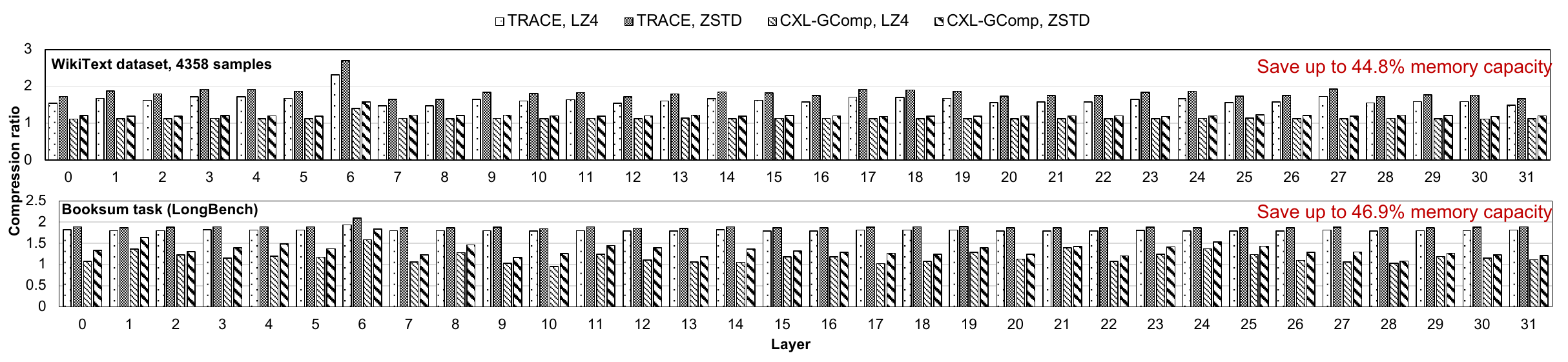}
	\caption{\textbf{KV cache lossless compression ratio by layer} for LLaMA~3.1~8B (32 layers) on WikiText and BookSum, using 4\,KB blocks with LZ4/ZSTD.
    \emph{TRACE}: KV preprocessing (cross-token channel grouping + exponent-delta de-correlation, Sec.~\ref{sec:compression}) on top of the bit-plane layout.
    \emph{CXL-GComp}: direct lossless compression on the standard token-major layout.}
\label{fig.kv cache by layer compression ratio}
\end{figure*}

\smallskip
\noindent\textbf{KV cache.}\quad
We profile BF16 KV blocks across all 32 layers of LLaMA~3.1~8B on WikiText~\cite{wikitext-website} and BookSum (LongBench)~\cite{kryscinski2021booksum}. Fig.~\ref{fig.kv cache by layer compression ratio} reports per-layer averages. Direct generic compression on the standard token-major KV stream remains weak: with ZSTD, CXL-GComp achieves overall ratios of 1.21 (WikiText) and 1.33 (BookSum). TRACE increases KV compressibility across essentially all layers, reaching 1.81 (WikiText) and 1.88 (BookSum), i.e., 44.8\% and 46.9\% footprint reduction. Relative to CXL-GComp, this corresponds to a 50.3\% (WikiText) and 41.7\% (BookSum) increase in lossless KV compression ratio under the same codec. The most compressible layers reach 2.69$\times$ (ZSTD) and 2.31$\times$ (LZ4) on WikiText, and 2.10$\times$ (ZSTD) and 1.93$\times$ (LZ4) on BookSum, consistent with Mechanism~I exposing low-entropy structure before bit-plane packing.

\smallskip
\noindent\textbf{Weights.}\quad
Table~\ref{tab:model weight compression ratio} reports TRACE’s lossless compression ratio on the device-side representation and the resulting footprint reductions. We separate (i) \emph{lossless savings}, the additional reduction from TRACE beyond the offline stored format, and (ii) \emph{total savings vs.\ BF16}, which combines an optional lossy quantization step (e.g., converting BF16 weights to FP8/INT4) with TRACE’s lossless compression on the resulting bitstream. BF16 weights show the largest incremental lossless gains (e.g., LLaMA~3.1~8B achieves 1.34$\times$ under ZSTD, i.e., 25.2\% lossless reduction). After aggressive quantization, the remaining lossless headroom is smaller but still measurable (e.g., INT4 adds 0.9--2.1\% lossless reduction on top of the 4$\times$ reduction from BF16 to INT4, yielding 75.2--75.5\% total savings vs.\ BF16 across models). Because TRACE operates on the device-side physical representation, these gains are orthogonal to quantizers such as GPTQ~\cite{frantar2022gptq}.

\begin{table}[htbp]
\centering
\caption{Lossless compression ratios on weights under TRACE, and total footprint reduction when combined with optional lossy quantization.}
\label{tab:model weight compression ratio}
\resizebox{\linewidth}{!}{%
\begin{tabular}{@{}lllll@{}}
\toprule
Model          & Precision & Comp. Ratio & Lossless Savings & Total Savings (vs.\ BF16) \\ \midrule
LLaMA 3.1 8B   & BF16      & 1.34              & 25.2\%          & 25.2\%      \\
               & FP8       & 1.09              & 8.3\%           & 54.1\%      \\
               & INT4      & 1.01              & 0.9\%           & 75.2\%      \\
LLaMA 3.1 70B  & BF16      & 1.34              & 25.6\%          & 25.6\%      \\
               & FP8       & 1.10              & 9.3\%           & 54.6\%      \\
               & INT4      & 1.02              & 2.1\%           & 75.5\%      \\
Mixtral 8$\times$7B   & BF16      & 1.32              & 24.4\%          & 24.4\%      \\
               & FP8       & 1.09              & 8.0\%           & 54.1\%      \\
               & INT4      & 1.01              & 1.2\%           & 75.3\%      \\
LLaMA MoE 3.5B & BF16      & 1.33              & 24.9\%          & 24.9\%      \\
               & FP8       & 1.11              & 9.9\%           & 54.9\%      \\
               & INT4      & 1.02              & 1.6\%           & 75.4\%      \\ \bottomrule
\end{tabular}%
}
\end{table}

\begin{figure}[htbp]
	\centering
    \includegraphics[width=\linewidth]{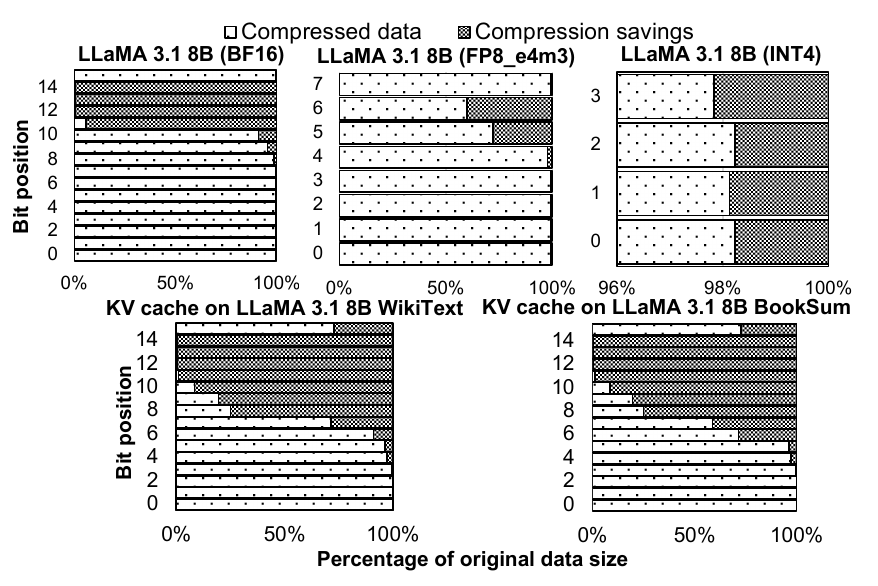}
	\caption{\textbf{Plane-level compressibility (ZSTD, 4\,KB blocks).} Per-plane compression ratios for BF16/FP8/INT4 weights and BF16 KV (LLaMA~3.1~8B on WikiText and BookSum). High-order exponent planes are consistently the most compressible, and KV exponent planes benefit further from TRACE’s channel-major grouping and exponent-delta normalization.}
\label{fig.model parameter compressibility}
\end{figure}

\smallskip
\noindent\textbf{Plane-level evidence.}\quad
Fig.~\ref{fig.model parameter compressibility} reports ZSTD compression ratios by bit-plane (4\,KB blocks). For BF16 weights, the most significant exponent planes dominate the overall reduction, matching the 1.34$\times$ end-to-end ratio in Table~\ref{tab:model weight compression ratio}; within a block, magnitudes cluster, so exponent fields take a small set of values that become long runs once stored as contiguous planes. After FP8/INT4 quantization, this redundancy is reduced by construction (fewer exponent/mantissa states and more packed representations), so the per-plane headroom narrows. KV exhibits the same exponent-first trend for a different reason. Mechanism~I normalizes exponent variation within channel-major groups, so exponent patterns repeat across tokens, explaining the large end-to-end KV reductions in Fig.~\ref{fig.kv cache by layer compression ratio} despite weak direct compression under CXL-GComp.

\subsection{DRAM Access Efficiency}
\label{sec:DRAM access efficiency}

This subsection isolates \emph{device-side DRAM traffic} under dynamic/elastic precision (Section~\ref{sec:runtime_structure}). We test whether lower effective precision translates into \emph{physical} savings inside the CXL device, i.e., fewer DRAM activations and fewer DRAM bytes moved. TRACE’s Mechanism~II (elastic precision access) achieves this by fetching only the required bit-planes for each precision view.

We focus on weights because they are reread every decoding step and dominate recurring reads at short-to-moderate contexts. Studying weights in isolation also avoids entanglement with KV tiering policies, so any change can be attributed to plane-aligned fetch.

To isolate plane-aligned fetch, we disable block compression in this subsection and compare word-fetch vs plane-fetch on the same uncompressed storage. 

We evaluate two elastic precision granularities: 
(i) \textit{per-expert} decisions in MoE-style control flow (LLaMA 3.1 8B/70B~\cite{dubey2024llama}, Mixtral~8$\times$7B~\cite{jiang2024mixtral}, and LLaMA-MoE-3.5B~\cite{zhu2024llama}), and 
(ii) \textit{per-head/per-neuron} decisions in OPT~30B~\cite{zhang2022opt}. All DRAM results use DRAMSim3~\cite{li2020dramsim3} with 4 channels per module and 10$\times$4 DDR5-4800 devices per channel.

\smallskip
\noindent\textbf{Average bits/weight.}\quad
We report results versus average bits/weight, the footprint-weighted mean effective bitwidth of weights actually fetched during decoding. The runtime assigns a discrete precision view per unit (expert/head/neuron) based on its importance, and averaging these choices over a decoding window yields the expected number of bit-planes fetched per accessed weight. We treat average bits/weight as a budget knob that controls how aggressively the device can reduce DRAM traffic.

\begin{figure*}[htbp]
	\centering
    \includegraphics[width=\linewidth]{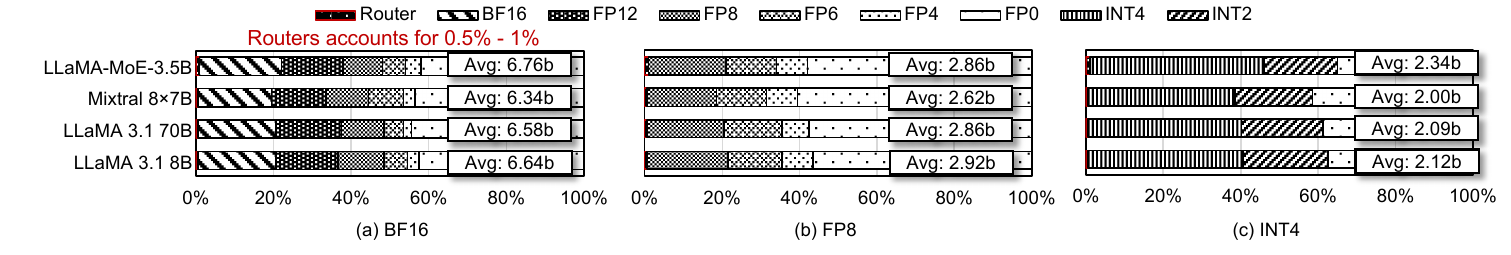}
	\caption{\textbf{Runtime precision distribution} of MoDE-controlled weights on WikiText-2 for LLaMA 3.1 8B/70B, Mixtral 8$\times$7B, and LLaMA-MoE-3.5B across BF16/FP8/INT4 bases (routers in BF16).}
\label{fig.precision_distribution}
\end{figure*}

\smallskip
\noindent\textbf{Granularity~I: per-expert.}\quad
We adapt LLaMA 3.1 8B/70B and Mixtral 8$\times$7B to a Mixture-of-Depth-and-Experts (MoDE) control flow~\cite{raposo2024mixture} that caps per-block precision and assigns per-expert precision within that cap. Dense MLPs in LLaMA are converted to MoE to expose expert-level control. We reduce calibration cost with LoRA~\cite{hu2021lora} on C4, and prepare FP8 and INT4 variants using AutoFP8~\cite{autofp8} and GPTQ~\cite{frantar2022gptq} on UltraChat~\cite{ding2023enhancing}. Fig.~\ref{fig.precision_distribution} summarizes the resulting runtime precision mixes on WikiText-2.

For clarity in the figures, we label TRACE as \textit{T} and the conventional byte/word-fetch baseline as \textit{B} (CXL-Plain). Beyond the figure labels, we refer to the two designs by name. CXL-Plain always fetches full fixed-width containers and applies any precision conversion after the read, whereas TRACE performs plane-aligned fetch and activates only the bit-planes required by the requested precision view.

\begin{figure}[htbp]
	\centering
    \includegraphics[width=\linewidth]{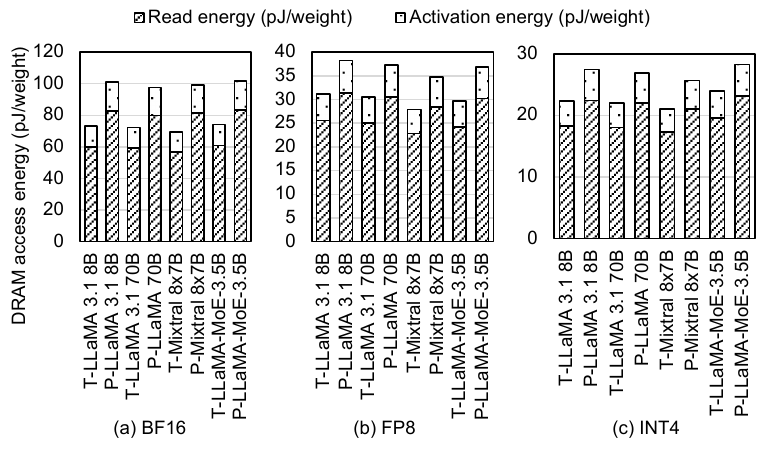}
	\caption{\textbf{DRAM access energy} (per-expert granularity)  under CXL-Plain (P) versus TRACE (T) across BF16/FP8/INT4 bases on WikiText-2.}
\label{fig.access_energy}
\end{figure}

Fig.~\ref{fig.access_energy} reports DRAM access energy for weight reads under per-expert elastic precision. TRACE reduces DRAM energy by up to 29.9\% on BF16 bases (27.8\% on LLaMA~3.1~8B, 25.9\% on LLaMA~3.1~70B, 29.9\% on Mixtral~8$\times$7B, and 27.2\% on LLaMA-MoE-3.5B). Savings persist for quantized bases but taper as intrinsic bitwidth shrinks (e.g., 19.6\% for FP8 and 17.9\% for INT4 on Mixtral), since fewer planes remain to skip once the offline format is already low precision.

\begin{figure}[htbp]
	\centering
    \includegraphics[width=\linewidth]{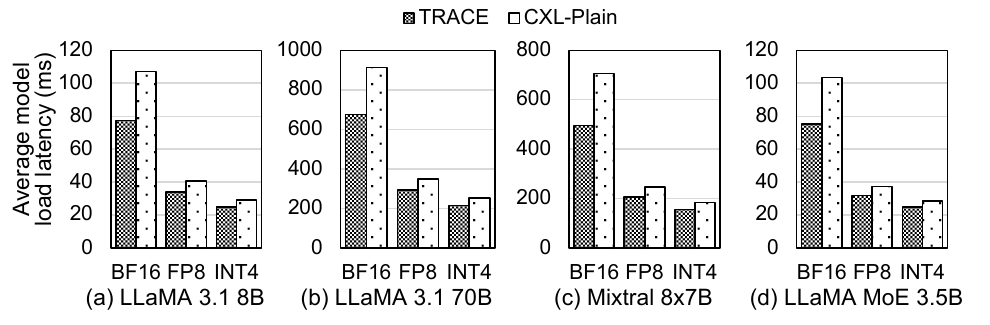}
	\caption{\textbf{Average model load latency} (per-expert granularity) under CXL-Plain (P) versus TRACE (T) for LLaMA~3.1~8B/70B, Mixtral~8$\times$7B, and LLaMA-MoE-3.5B across BF16/FP8/INT4 bases on WikiText-2.}
\label{fig.model load latency}
\end{figure}

Fig.~\ref{fig.model load latency} reports model load latency for weights. We define \emph{model load latency} as the device-side DRAM service time per decoding step for the weight reads, excluding CXL link transfer, host compute, and KV traffic. We report average latency over many decoding steps because routing and precision selection change across tokens, which changes how many bit-planes are fetched and the resulting DRAM access pattern.
TRACE reduces load latency by up to 30.0\% on BF16 bases. For example, Mixtral~8$\times$7B decreases from 705.90~ms to 495.06~ms (30.0\%), and LLaMA~3.1~70B decreases from 910.58~ms to 674.73~ms (25.9\%). Quantized variants also benefit: for LLaMA~3.1~70B, FP8 decreases from 348.65~ms to 293.27~ms and INT4 decreases from 251.03~ms to 214.11~ms.

\smallskip
\noindent\textbf{Granularity~II: per-attention-head and per-neuron.}\quad
To validate finer-grained control, we reuse the OPT~30B setup where runtime precision is selected per attention head and per MLP neuron. We treat all weights in one head or one neuron as a chunk, aligned to TRACE’s plane stripes so that the device fetch remains proportional to the requested precision view. Each head contains $3.7\times 10^6$ weights and each MLP neuron contains $7.2\times 10^3$ weights. We compare CXL-Plain (byte/word fetch that always reads full containers) against TRACE (plane-aligned fetch that activates only the required bit-planes).

\begin{figure}[htbp]
\centering
\includegraphics[width=\linewidth]{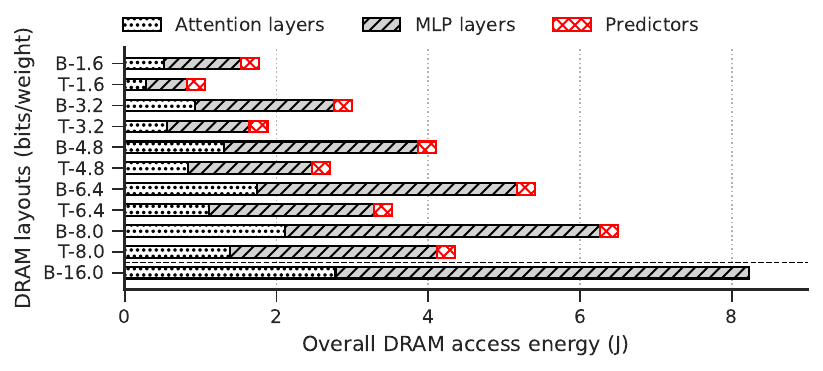}
\caption{\textbf{Total DRAM access energy for one full model load} (per-head and per-neuron granularity (OPT~30B)).  \textbf{CXL-Plain} (B) uses byte/word fetch, \textbf{TRACE} (T) uses plane-aligned fetch. Bars sweep the average bits/weight target. \texttt{B-16.0} shows the full 16-bit load.}
\label{fig.overall DRAM access energy}
\end{figure}

\begin{figure}[htbp]
\centering
\includegraphics[width=\linewidth]{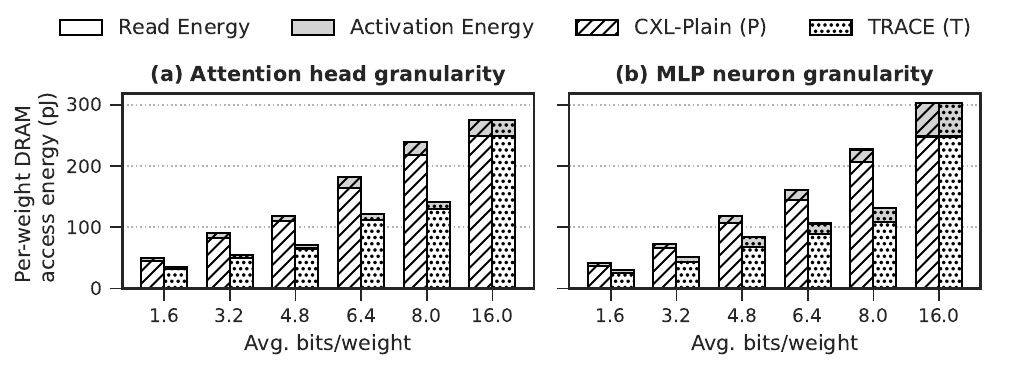}
\caption{\textbf{Per-head and per-neuron granularity (OPT~30B):} Per-weight DRAM access energy under elastic precision, comparing \textbf{CXL-Plain} (P) and \textbf{TRACE} (T) at (a) head granularity and (b) neuron granularity. Stacked bars separate read and activation energy across bits/weight targets.}
\label{fig.per weight DRAM energy}
\end{figure}

Fig.~\ref{fig.overall DRAM access energy} reports total DRAM access energy for loading the full model once. TRACE reduces total DRAM access energy by up to 40.3\% relative to CXL-Plain. Fig.~\ref{fig.per weight DRAM energy} breaks this down per weight. For attention heads at 1.6/4.8/8.0~bits/weight targets, CXL-Plain costs 49.6/118.9/238.9~pJ per weight, while TRACE costs 34.5/70.8/141.2~pJ, corresponding to 30.5\%, 40.4\%, and 40.9\% reductions. For MLP neurons, TRACE lowers per-weight energy by 19.4\%, 20.3\%, and 33.9\% at the same targets. Latency follows the same trend. 

\subsection{Hardware Implementation}
\label{sec:ppa_latency}

We implement TRACE in SystemVerilog following Fig.~\ref{fig:microarchitecture}: alias decoding and plane-mask generation, hierarchical metadata lookup with an SRAM-resident plane-index cache, a multi-lane LZ4 codec with transpose/reconstruction, and a plane-aware DDR scheduler. We also implement \textit{CXL-Plain}, as the conventional Type-3 controller with word-major fixed-width DRAM fetch, and \textit{CXL-GComp}, which augments CXL-Plain with an inline LZ4 codec and staging buffers. TRACE builds on CXL-GComp by adding the bit-plane substrate (plane indexing and masking) and plane-aware scheduling so reduced-precision views issue plane-granular DRAM reads, while preserving an unmodified host-visible CXL.mem interface. The microarchitecture is provisioned to match the device DDR budget and is designed to sustain 256\,GB/s device-side throughput. All designs are synthesized in ASAP7 7\,nm~\cite{clark2016asap7} at 2\,GHz and 0.7\,V.

\smallskip
\noindent\textbf{PPA.}\quad
Table~\ref{tab:ppa} reports area, power, and load-to-use latency. Relative to CXL-GComp, TRACE increases total area from 6.66 to 7.14\,mm\textsuperscript{2} (+0.48\,mm\textsuperscript{2}, 7.2\%) and power from 21.4 to 22.4\,W (+1.0\,W, 4.7\%). The increase is dominated by the metadata subsystem: TRACE expands the on-chip metadata from 0.42 to 0.83\,mm\textsuperscript{2} (+0.41\,mm\textsuperscript{2}) to cache plane indices and avoid DRAM round-trips for block and plane pointer resolution. The remaining area comes from plane support logic, including the transpose/reconstruction block (0.06\,mm\textsuperscript{2}) and a slightly larger scheduler (0.02 to 0.03\,mm\textsuperscript{2}) to track plane-level request constraints. Importantly, the codec datapath (1.92\,mm\textsuperscript{2}) and staging SRAM (0.62\,mm\textsuperscript{2}) are identical between TRACE and CXL-GComp. TRACE reuses the same 32-lane LZ4 engine and buffering, and only changes what the engine is asked to fetch and reconstruct.

\begin{table}[htbp]
\centering
\caption{Hardware cost (ASAP7 7\,nm @ 2\,GHz, 0.7\,V). CXL-GComp is the primary baseline.}
\label{tab:ppa}
\small
\setlength{\tabcolsep}{4pt}
\resizebox{.8\linewidth}{!}{%
\begin{tabular}{lccc}
\toprule
 & \textbf{CXL-Plain} & \textbf{CXL-GComp} & \textbf{TRACE} \\
\midrule
Area (mm\textsuperscript{2}) & 3.91 & 6.66 & 7.14 \\
Power (W) & 9.0 & 21.4 & 22.4 \\
Load-to-use (cycles) & 71 & 84 & 89 \\
\midrule
\multicolumn{4}{l}{\textit{Area breakdown (mm\textsuperscript{2}):}} \\
PHY & 3.50 & 3.50 & 3.50 \\
Codec & -- & 1.92 & 1.92 \\
Codec SRAM & -- & 0.62 & 0.62 \\
Metadata & 0.21 & 0.42 & 0.83 \\
Scheduler & 0.02 & 0.02 & 0.03 \\
Transpose/Recon. & -- & -- & 0.06 \\
Other & 0.18 & 0.18 & 0.18 \\
\bottomrule
\end{tabular}}
\vspace{-2mm}
\end{table}

The load-to-use service time increases from 84 to 89 cycles (+5 cycles, 6.0\%, 2.5\,ns at 2\,GHz). This delta comes from additional front-end translation (alias decode and plane-mask generation) and plane-aware scheduling, while metadata resolution remains fast due to the plane-index cache. The result is that TRACE achieves plane-granular fetch with single-request latency close to a generic-compressed controller, and the system-level gains are driven by reduced bytes moved rather than by trading off higher controller latency.

\begin{figure}[htbp]
    \centering
    \includegraphics[width=\linewidth]{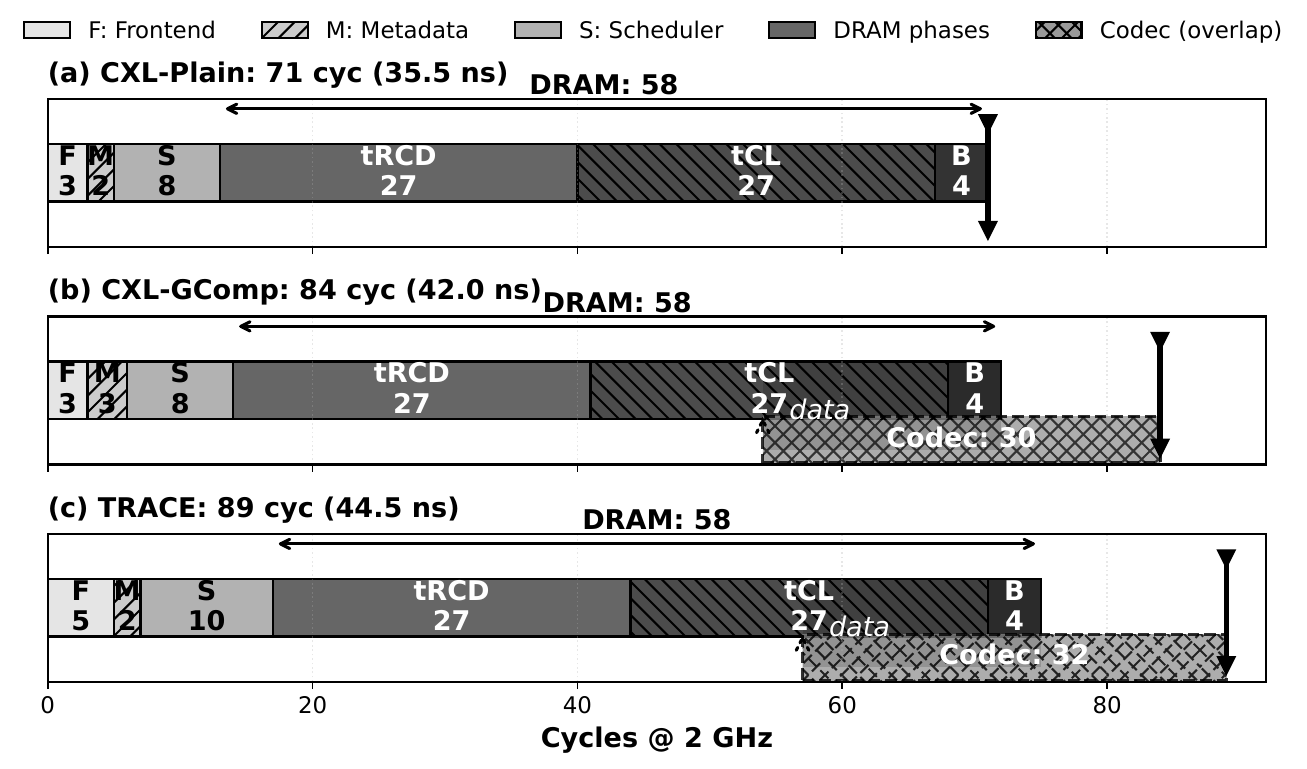}
    \caption{\textbf{Pipeline timing breakdown (metadata-cache hit).}
    The codec is streaming and overlaps with the DRAM access window; the exposed critical path is dominated by metadata lookup and plane scheduling.
    Stages: F (frontend: CXL.mem decode), M (metadata: address translation/compression indices), S (scheduler: DDR arbitration), tRCD (row activation), tCL (column access), B (burst transfer).
    The dashed codec bar shows overlap with DRAM. The downward arrow marks the critical path endpoint.}
    \label{fig:pipeline_timing}
\end{figure}

\smallskip
\noindent\textbf{Latency.}\quad
Fig.~\ref{fig:pipeline_timing} reports the controller load-to-use service time, decomposed into front-end translation, metadata resolution, DDR scheduling, and the DRAM access window. Here, load-to-use is device-local (controller front-end to data ready for return) and excludes host/link/fabric and system-level queuing. The breakdown assumes a metadata-cache hit in the on-chip plane-index cache. On a metadata-cache miss, the controller performs an additional DRAM read to fetch the plane-index entry from the device-resident metadata store. This adds roughly one extra DRAM access window (tRCD+tCL+burst) before the data-plane reads, but does not require rereading the data block.

CXL-Plain completes in 71 cycles (35.5\,ns): 3 cycles front-end decode, 2 cycles metadata translation, 8 cycles scheduling, and a 58-cycle DRAM window.

CXL-GComp completes in 84 cycles (42.0\,ns), adding 13 cycles (18.3\%) over CXL-Plain. The exposed delta comes from locating variable-length blocks and codec bookkeeping in the controller. The LZ4 datapath is streaming and overlaps with the DRAM window, so decompression does not sit on the critical path.
TRACE completes in 89 cycles (44.5\,ns), adding 5 cycles (6.0\%) over CXL-GComp. This overhead comes from alias/plane-mask generation in the front-end (5 vs.\ 3 cycles) and plane-aware scheduling (10 vs.\ 8 cycles). Metadata resolution remains 2 cycles due to the on-chip plane-index cache. Transpose/reconstruction is also streamed and overlapped with the DRAM window.
In total, TRACE adds 9\,ns (25.4\%) over CXL-Plain while keeping latency close to a generic-compression controller.

\begin{figure}[htbp]
    \centering
    \includegraphics[width=\linewidth]{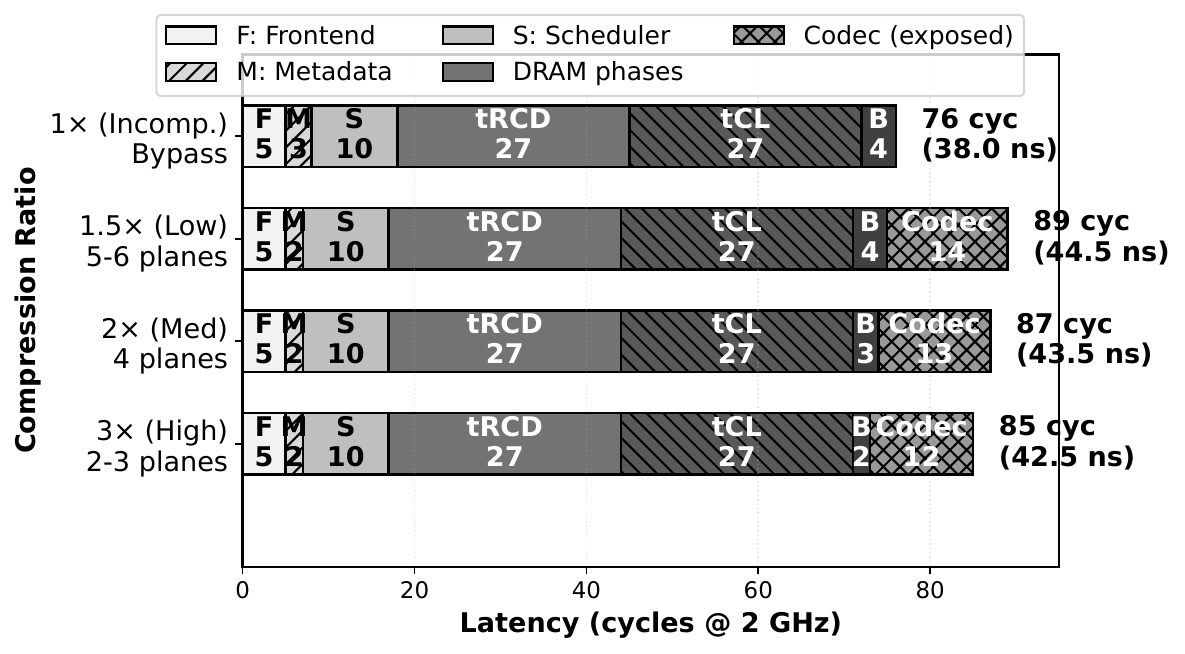}
\caption{\textbf{TRACE latency vs. compression ratio (metadata-cache hit).}
    Stages: F (frontend), M (metadata), S (scheduler), tRCD/tCL/B (DRAM phases),
    Codec (exposed).
    Higher compression reduces both burst time (fewer bit-planes) and codec latency
    (less data), improving from 89 cycles (1.5×) to 85 cycles (3×).
    Incompressible data bypasses codec, achieving 76 cycles.}
\label{fig:trace_compression_latency}
\end{figure}

Fig.~\ref{fig:trace_compression_latency} shows how TRACE’s load-to-use latency varies with compression ratio under the streaming codec pipeline. Higher compression fetches fewer planes, shortening the DRAM burst and reducing exposed codec work, so total service time drops from 89 cycles (44.5\,ns) at 1.5$\times$ to 85 cycles (42.5\,ns) at 3$\times$. Incompressible blocks take the bypass path and complete in 76 cycles (38.0\,ns), since the controller skips codec processing and returns the raw planes with only fixed control overhead.

\section{Related Work}
\label{sec:related work}

\noindent\textbf{Memory compression in systems.}\quad
Hardware-assisted memory compression has been studied for decades, including line-oriented designs and metadata-efficient schemes for reducing DRAM traffic and footprint~\cite{zhao2015buri}. Real systems also rely heavily on block compression in memory-resident data platforms and key-value stores to lower effective memory cost and bandwidth pressure~\cite{dageville2016snowflake,xie2024zipcache,lahiri2015oracle}.

\smallskip
\noindent\textbf{Reducing LLM footprint and dynamic execution.}\quad
LLM efficiency work primarily reduces bytes via lossy compression (quantization) and structured sparsity. Post-training quantization methods such as GPTQ and AWQ lower weight precision~\cite{frantar2022gptq,lin2024awq}, while SparseGPT combines sparsity and quantization to cut compute and memory traffic~\cite{frantar2023sparsegpt}. Separately, input-dependent systems exploit contextual importance and conditional computation (e.g., Deja Vu~\cite{liu2023deja}, MoE/MoD~\cite{raposo2024mixture}) to concentrate resources on the most relevant parts of the model. For long-context decoding, KV cache management (e.g., page selection/eviction in Quest~\cite{tang2024quest}) and lossy KV quantization reduce the active KV working set and its effective precision. TRACE targets the complementary, lossless reduction of bytes moved by the capacity tier via device-internal layout transformation and compression, and it can be combined with these lossy runtime policies. Evaluating such combined policies is a natural future direction.

\smallskip
\noindent\textbf{Bit-plane layouts and precision-scalable retrieval.}\quad
Bit-plane organization is a known way to expose bit-level redundancy for compression and to support precision-scalable fetch. BPC demonstrates improved compressibility from plane regrouping~\cite{kim2016bit}. More recent work uses plane-based layouts to enable any-precision execution or context-aware retrieval (e.g., AnyBCQ~\cite{park2024any}, SmartQuant~\cite{xie2024smartquant}). TRACE builds on this general principle in the CXL Type-3 setting: for any model state stored as tensors (e.g., Mamba~\cite{gu2024mamba}), since adjacent elements exhibit redundancy (e.g., spatial locality in value distributions or repeated high-order fields), a bit-plane layout can expose low-entropy streams to commodity lossless codecs and enable precision-proportional fetch by physically skipping unneeded planes.

\section{Conclusion}
TRACE reduces CXL-tier traffic for LLM inference by changing the device-internal representation while preserving the standard CXL.mem load/store interface. TRACE stores weights and KV in a channel-major, bit-plane layout that (i) converts mixed-field word streams into low-entropy plane streams that commodity lossless codecs can compress effectively and (ii) enables precision views that physically skip unneeded bit-planes, making device-side DRAM work proportional to requested precision. Across public models, TRACE reduces BF16 weight footprint by 25.2\% and BF16 KV footprint by 46.9\% with no accuracy loss, with per-layer KV ratios peaking at 2.69$\times$. In trace-driven system modeling, these savings translate into large long-context gains once KV spills to CXL, including a 4.24$\times$ throughput improvement at 128k tokens on GPT-OSS-120B-MXFP4. Under elastic precision control, plane-aligned fetch reduces weight-read DRAM access energy by up to 40.3\% and reduces model-load latency by up to 30.0\% depending on granularity. A SystemVerilog implementation in ASAP7 7\,nm indicates that these gains come at modest controller cost relative to a generic inline-compression baseline: +7.2\% area, +4.7\% power, and +6.0\% load-to-use latency at 2\,GHz and 0.7\,V.

\bibliographystyle{IEEEtran}
\bibliography{reference}


\vspace{-30pt}
\begin{IEEEbiographynophoto}{Rui Xie}
is a Ph.D. candidate in the Department of Electrical, Computer, and Systems Engineering at Rensselaer Polytechnic Institute, Troy, NY, USA. His research focuses on efficient memory system design for Large Language Models. He received the B.E. degree in Microelectronics and Science and Engineering from Southern University of Science and Technology, China.
\end{IEEEbiographynophoto}
\vspace{-30pt} 
\begin{IEEEbiographynophoto}{Asad Ul Haq}
is currently pursuing the Ph.D. degree in the Department of Electrical, Computer, and Systems Engineering at Rensselaer Polytechnic Institute, Troy, NY, USA. His research interest is in computational CXL memory. He received the B.S. degree in Electrical Engineering from the National University of Sciences and Technology, Pakistan.
\end{IEEEbiographynophoto}
\vspace{-30pt} 
\begin{IEEEbiographynophoto}{Yunhua Fang}
is currently pursuing the Ph.D. degree in the Department of Electrical, Computer, and Systems Engineering at Rensselaer Polytechnic Institute, Troy, NY, USA. His research focuses on memory system design for AI-centric infrastructure. He received his B.S. degree in Computer Science from the University of California, Davis.
\end{IEEEbiographynophoto}
\vspace{-30pt} 
\begin{IEEEbiographynophoto}{Linsen Ma}
received the Ph.D. degree in 2025 from the Department of Electrical, Computer, and Systems Engineering at Rensselaer Polytechnic Institute, Troy, NY, USA. His doctoral research focused on efficient data management over computational storage. He received the M.S. and B.S. degree in Electrical Engineering from Rensselaer Polytechnic Institute.
\end{IEEEbiographynophoto}
\vspace{-30pt} 
\begin{IEEEbiographynophoto}{Zirak Burzin Engineer}
is a student at Wiseburn Da Vinci Science High School, El Segundo, CA, USA. His interests include computer architecture and machine learning. He contributed to this research through a summer program focused on high-performance computing.
\end{IEEEbiographynophoto}
\vspace{-30pt} 
\begin{IEEEbiographynophoto}{Liu Liu}
is an Assistant Professor in the Department of Electrical, Computer, and Systems Engineering at Rensselaer Polytechnic Institute, Troy, NY, USA. His research interests include elastic AI computing systems and architecture design. He received the B.S. degree from the University of Electronic Science and Technology of China, the M.S. degree in electrical and computer engineering, and the Ph.D. degree in computer science, both from the University of California, Santa Barbara.
\end{IEEEbiographynophoto}
\vspace{-30pt} 
\begin{IEEEbiographynophoto}{Tong Zhang}
(Fellow, IEEE) is a Professor in the Department of Electrical, Computer, and Systems Engineering at Rensselaer Polytechnic Institute, Troy, NY, USA. His current research areas are computer system design with the focus on memory-centric computing for AI and data science. He received the B.S. and M.S. degrees in electrical engineering from Xi'an Jiaotong University, China, and the Ph.D. degree in electrical and computer engineering from the University of Minnesota, Minneapolis.
\end{IEEEbiographynophoto}

\vfill

\end{document}